\documentclass[journal=acs photon.,manuscript=article]{achemso}
\usepackage{bm,physics,amsmath,amssymb,
	graphicx,xcolor,embedfile,comment,mathrsfs,mathtools,xfrac}
\usepackage[utf8]{inputenc}
\usepackage[normalem]{ulem}
\usepackage{subfigure}

\usepackage[sort&compress,numbers,super]{natbib}
\setkeys{acs}{doi = true}

\usepackage[T1]{fontenc}
\usepackage{tikz}

\graphicspath{{figures/}}

\makeindex

\title{Observation of Analogue Dynamic Schwinger Effect and Non-Perturbative Light Sensing in Lead Halide Perovskites}

\author{Dusan Lorenc}
\affiliation{ISTA (Institute of Science and Technology Austria),
    Am Campus 1, 3400 Klosterneuburg, Austria}

\author{Artem G. Volosniev}
\email{artem@phys.au.dk}
\affiliation{Department of Physics and Astronomy, Aarhus University, Ny Munkegade 120, DK-8000 Aarhus, Denmark}

\author{Ayan A. Zhumekenov}
\affiliation{KAUST (King Abdullah University of Science and Technology),
    Thuwal 23955, Saudi Arabia}
    \altaffiliation{Present address: School of Materials Science and Engineering (MSE), Nanyang Technological University, Singapore 639798}
    
\author{Seungho Lee}
\affiliation{ISTA (Institute of Science and Technology Austria),
    Am Campus 1, 3400 Klosterneuburg, Austria}  

\author{Maria Ib{\' a}{\~ n}ez}
\affiliation{ISTA (Institute of Science and Technology Austria),
    Am Campus 1, 3400 Klosterneuburg, Austria}  
    
 \author{Osman M. Bakr}
\affiliation{KAUST (King Abdullah University of Science and Technology),
    Thuwal 23955, Saudi Arabia} 
    
     \author{Mikhail Lemeshko}
\affiliation{ISTA (Institute of Science and Technology Austria),
    Am Campus 1, 3400 Klosterneuburg, Austria}   
    
\author{Zhanybek Alpichshev}
\email{alpishev@ist.ac.at}
\affiliation{ISTA (Institute of Science and Technology Austria),
    Am Campus 1, 3400 Klosterneuburg, Austria}

\begin{document}

\begin{abstract}

Dielectric breakdown of physical vacuum (Schwinger effect) is the textbook demonstration of compatibility of Relativity and Quantum theory. {\color{black} Although observing this effect is still practically unachievable, its analogue generalizations have been shown to be more readily attainable.} This paper demonstrates that a gapped Dirac semiconductor, {\color{black}methylammonium lead-bromide perovskite (MAPbBr$_3$)}, exhibits analogue dynamical Schwinger effect. Tunneling ionization under deep sub-gap mid-infrared irradiation leads to intense photoluminescence in the visible range, in full agreement with quasi-adiabatic theory. {\color{black} In addition to revealing a gapped extended system suitable for studying the {\color{black} analogue} Schwinger effect, this observation holds great potential for non-perturbative field sensing, i.e., sensing electric fields through non-perturbative light-matter interactions. First, this paper illustrates this by measuring the local deviation from the nominally cubic phase of a perovskite single crystal, which can be interpreted in terms of frozen-in fields.} Next, it is shown that {\color{black} analogue} dynamic Schwinger effect can be used for nonperturbative amplification of {\color{black} non-parametric upconversion process} in perovskites driven simultaneously by multiple optical fields. This discovery demonstrates the potential for material response beyond perturbation theory in the Schwinger regime, offering extremely sensitive light detection and amplification across an ultrabroad spectral range not accessible by conventional devices.
\end{abstract}

    \textbf{Keywords:} Photoluminescence; Quasi-adiabatic tunneling ionization; Dirac equation; Landau-Dykhne approximation; Electric field detection

\section{Introduction}
One of the most important insights coming from the synthesis of Quantum Mechanics and Special Relativity is the realization that vacuum is not empty. Instead, it is rather a fluctuating sea of virtual particle-antiparticle pairs, which can be in principle made real in the presence of external fields~\cite{Schwinger1951}. However, this process is negligible unless the fields are of the order of $E_S \approx 10^{18}$V/m at which point particle-antiparticle creation leads to the dielectric breakdown of vacuum.
  In view of the enormity of $E_S$ and its intuitive interpretation as the maximal static electric field attainable in principle, it is natural that the actual attempts to realize vacuum breakdown (“Schwinger Effect”, SE) focus on its dynamical versions where the vacuum pair creation is achieved under the influence of time-dependent fields~\cite{Mourou2006},    {\color{black} which are easier to create (from a modern technological point of view) than the static ones.  To further facilitate the observation of SE, theoretical studies suggest either to increase the frequency of the driving field~\cite{Brezin1970,Popov1971} or to tailor the driving field profiles~\cite{Ringwald2001,Schtzhold2008}}. However, any theoretical proposal must rely on certain simplifying assumptions, which due to the inherently non-perturbative character of SE may be decisive to the outcome~\cite{Gies2005,Linder2015,Torgrimsson2017}. Given that current technology is not capable to test these ideas in practice, it is hard to tell at the moment how realistic will they ultimately turn out to be. 

{\color{black} In this context it appears desirable to be able to simulate the Schwinger effect and the various approaches to it in more accessible settings, such as cold-atom quantum simulators~\cite{Szpak2012,Kasper2016,Pieiro2019} or semiconductors~\cite{Zawadzki2005,Allor2008,FillionGourdeau2015,Linder2018}. As the connection between the Landau-Zener transition~\cite{Sauter1931,Zener1934} and the Schwinger effect is well understood (see, e.g., Ref.~\cite{Cohen2008}), arguably, the most theoretically explored simulator is a semiconductor with a gapped Dirac-like band dispersion~\cite{Zawadzki2005,Linder2018}}. It is straightforward to demonstrate then that the threshold field in this case would be $E_s \sim \Delta/ea \sim 10^8$V/m, which is readily achievable in modern laboratories (here $\Delta$, $a$ and $e$ are the band gap, unit lattice length and the elementary charge, respectively). However, it turns out that the observation of this scaled-down version of SE also poses significant challenges. The main difficulty here lies in the fact that despite an abundance of far-reaching analogies~\cite{Wehling2014,Zawadzki2017}, there is an important difference between the ``true'' Dirac field in vacuum and a Dirac material in that the former features relativistic invariance while the latter does not. The important consequence in question is that a free particle in vacuum can in principle be accelerated indefinitely, while a charge carrier in a semiconductor can produce secondary particles once it acquires a certain threshold kinetic energy comparable to the band-gap~\cite{Jena2022}. Since the threshold for this process $E_{\mathrm{av}} \sim \Delta/e\lambda_{\mathrm{mfp}}$ is significantly lower than that for tunneling ionization (here $\lambda_{\mathrm{mfp}} > a$ is the mean free path of charged carriers in the material), the population of secondary charge carriers will tend to dominate over tunnel-ionized carriers, often ending up in a catastrophic avalanche breakdown of the material. As a result of these complications, the observations of tunneling ionization have been limited so far either to the cases where the relevant tunneling region is restricted to a narrow part of space to avoid avalanche breakdown {\color{black}(e.g. tunneling ionization in single atoms \cite{Raizer1991}, in a spatially localized graphene junction~\cite{Schmitt2023} or tunnel diodes \cite{Esaki1958} and tunnel junctions \cite{Giaever1961, Chen2007} in electronics); or to extended systems with no gap \cite{Berdyugin2022}. 
However, to draw a direct analogy with high-energy physics, it is essential to identify materials and conditions that enable the realization of a Schwinger-like effect in an {\it extended} Dirac material with a {\it finite} bandgap.} {\color{black}[Note that in this work, `Dirac material' refers to a system whose underlying Hamiltonian exhibits a Dirac-like spin-momentum coupling and a particle-antiparticle structure. This definition, which is standard nowadays~\cite{Wehling2014}, refines early works based upon a Dirac-like energy-momentum relation~\cite{Zawadzki2017}.]}

In this Article, we report the observation of dynamical tunneling ionization in a single-crystal sample of lead halide perovskite (LHP), which realizes a gapped Dirac system~\cite{Jin2012}. 
In order to suppress avalanche formation and emphasize tunneling, we note that the former needs a finite amount of time to build up while the latter occurs quasi-instantaneously. Therefore, we seek to induce ionization with alternating electric fields using frequencies that, on the one hand, correspond to photon energies significantly smaller than the band gap $\hbar \omega \ll \Delta$ (to ensure that we operate in the quasi-adiabatic regime); on the other hand, they should oscillate fast enough to avoid an avalanche breakdown ($\omega \gtrsim e E /\sqrt{m \Delta}$, where $E$ and $m$ are the magnitude of the applied electric field and the band mass of the charge carrier, respectively). Additionally, we are careful to keep our irradiation intensity levels
well below impact ionization thresholds for our pulse durations $\tau \approx 300$fs to suppress the population of secondary charge carriers~\cite{Rethfeld2017}. Operation with relatively small intensities also allows us to avoid `memory' effects in our system, which are extrinsic to tunneling ionization (see Supporting Information). This is necessary to reveal the tunneling processes inherent in the material and also to be able to describe them using quasi-adiabatic methods. 

\begin{figure*}
\includegraphics[scale=0.5]{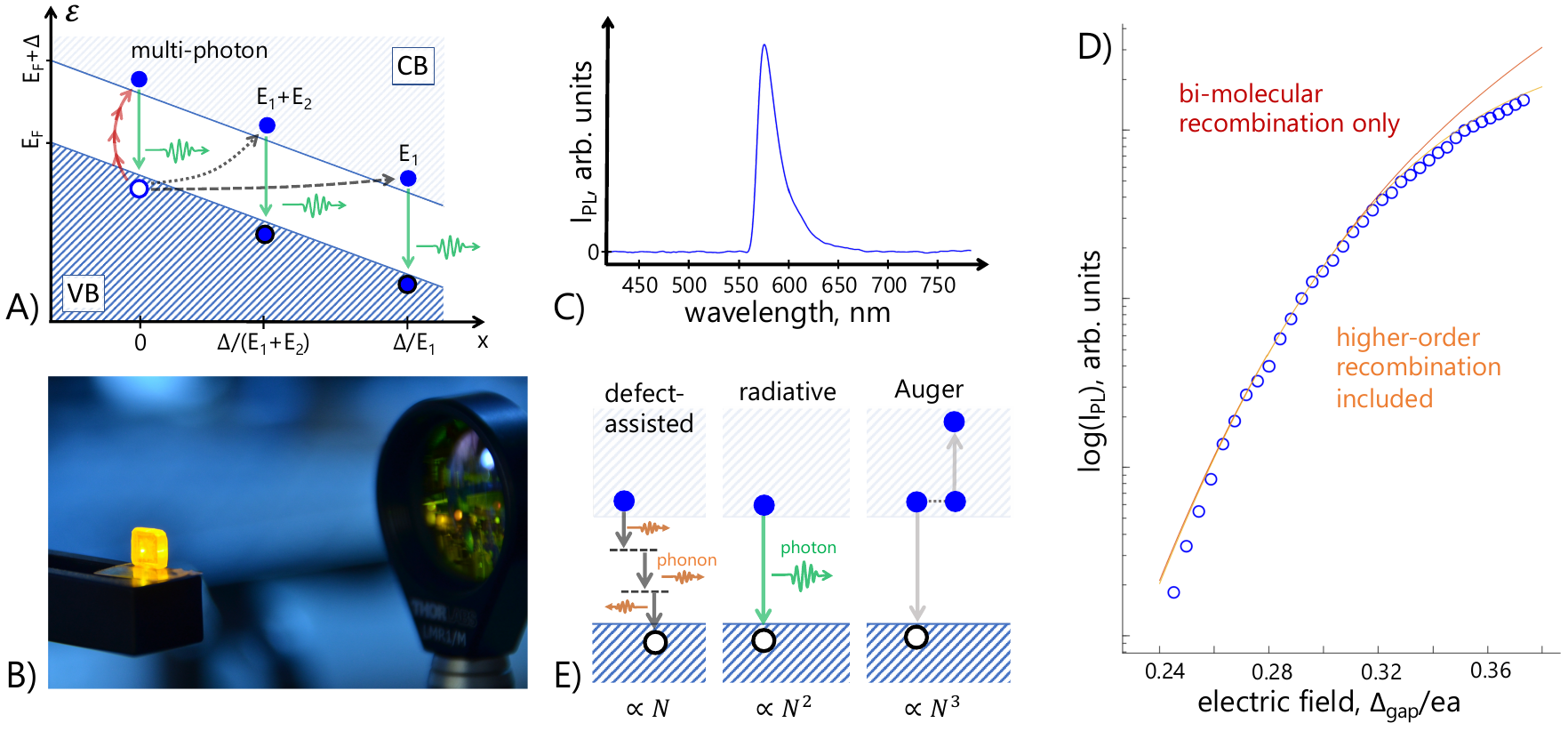}
\caption{{\color{black}A) Schematic diagram of ionization across the energy gap $\Delta$ due to multiphoton process, and due to tunneling under the influence of different applied field amplitudes ($E_1$ and $E_1+ E_2$). It can be seen that the magnitude of $E$ determines the width of the forbidden range, thus affecting the net tunneling rate exponentially (see the text for details); B) MAPbBr$_3$ crystal with PL coming from the bulk of the sample; C) PL spectrum of MAPbBr$_3$ single crystal sample pumped by $\lambda=4\mu$m radiation; D) PL spectra of MAPbBr$_3$ single crystal sample under $\lambda=4\mu$m pumping, together with our theoretical prediction for weak and intermediate electric fields (see the main text and Methods for details);  
E) An illustration of several common channels of photo-carrier recombination with characteristic rate dependencies on the density of charge carriers, $n$.}}
\label{fig:fig1}
\end{figure*}

Low irradiation intensities, however, imply that all measurements need to be performed in the regime where the concentration of tunnel-induced carriers is so low that standard transport measurements cannot be used for studying tunnel ionization. Therefore, our approach here is to infer the population of photo-excited carriers from the intensity of photoluminescence (PL) emitted upon their recombination with each other. The success of this approach crucially depends on two factors. First, the energy gap of the material should be free of defects to enhance the probability of tunneling directly into the conduction band. Second, the parent material should enjoy high photoluminescence quantum efficiency, namely the ratio of radiative recombinations to the total number of excitations~\cite{Braslavsky2007}. These considerations are the primary motivation for the choice of lead halide perovskites as the base for realizing the solid-state version of SE~\cite{Kirchartz2020}. Furthermore, effective behavior of lead halide perovskites renowned for excellent quantum optoelectronic properties~\cite{Goetz2020} is described very accurately by a Dirac equation with a non-zero mass~\cite{Becker2018,Volosniev2023, Volosniev2023a}, strengthening the analogy between our findings and the physics of SE.

{\color{black} Beyond fueling interest in basic strong-field physics discussed at the beginning of the paper, the observed tunneling ionization offers a promising avenue for sensitive electric field detection, even in the regimes where conventional sensors encounter limitations. To explore this avenue, we leverage the exponential dependence of tunneling ionization on electric field strength in the second part of the paper. Lead-halide perovskites appear as advantageous sensing materials, as the key characteristics of tunneling ionization remain robust with respect to variations in chemical/physical composition of the sample. To substantiate this claim, we will briefly discuss an alternative lead-halide perovskite, CsPbBr$_3$ (see Supporting Information and Ref.~\cite{Zhang2024}), as well as dust-like perovskite samples.
 }

{\color{black}
\section{Observation of tunneling ionization in MAP\lowercase{b}B\lowercase{r}$_3$} 
}
Lead-halide perovskites are known to feature strong sub-bandgap response in terms of two- and three-photon absorption~\cite{Zhou2021,Mei2022} whose analysis has been mainly focused around the photophysical properties of LHPs in the near-infrared regime as it can shed light onto surprising solar cell (photovoltaic) performance of LHPs~\cite{Jena2019}. At the same time, the response of the system to photons with lower frequency remains largely unexplored. In Fig.~\ref{fig:fig1}C we show the PL spectrum of {\color{black} single-crystal} MAPbBr$_3$ illuminated with mid-infrared radiation ($\lambda \approx 4\mu$m), see Experimental section for further details. Given the wavelength of PL ($\lambda_{\mathrm{PL}}\approx 570nm$) which roughly corresponds to the bandgap energy in MAPbBr$_3$ ($\Delta \approx 2.4$eV \cite{Saidaminov2015,Tilchin2016}) this process corresponds to a whopping frequency conversion factor of $\sim 10$ at the face value. Combined with the exponential sensitivity of this process to the intensity of the irradiating field in Fig.~\ref{fig:fig1}D, we are compelled to rationalize the observed PL under deep sub-gap irradiation as quasi-adiabatic tunnel ionization of electrons across the semiconductor band gap using the standard terminology of strong-field phenomena~\cite{Kruchinin2018} (similar behavior was observed in CsPbBr$_3$, which indicates the general character of the phenomenon; see Supporting Information and Ref.~\cite{Zhang2024}). This observation comprises the central finding of this work. In the following we will further substantiate {\color{black} this qualitative observation by a comprehensive quantitative analysis} of ionization in a periodically driven Dirac-like band structure.

\subsection{Quasi-adiabatic tunneling}
Tunneling is the quintessential quantum effect whereby a system undergoes a transition that can be described semi-classically by a trajectory of which at least a part passes through a classically-forbidden region in the parameter space. As a result the rate of this transition is strongly suppressed, being exponentially sensitive to the width of the classically-forbidden region~\cite{Landau1981}, see Fig.~\ref{fig:fig1}A. For instance, in  a static electric field, $E$, the rate of tunneling ionization $W$ across the gap $\Delta$ can be shown to be 
\begin{equation}
W \propto \exp\left( - E_s/|E|  \right) 
\label{eq:schwinger}
\end{equation}

\noindent with the characteristic cut-off field introduced above $E_s \sim \Delta/ea$ \cite{Sauter1931, Zener1934, Schwinger1951}. {\color{black} In the original Schwinger work~\cite{Schwinger1951}, the energy and length scales appear from the Dirac equation: $2 m_ec^2$ as the gap and the reduced Compton wavelength $\lambda_C=\hbar/(m_e c)$ in place of $a$, which leads to $E_S=m_e c^2/(e\lambda_C)$; here $m_e$ is the mass of an electron and $c$ is the speed of light. The huge disparity between these energy scales and those typical in semiconductor physics brings the observation of the analogue Schwinger effect significantly closer to experimental realization.} 

A convenient way to extend the expression in Eq.~(\ref{eq:schwinger}) to the problem of tunneling excitation of a semiconductor under the influence of time-dependent fields varying at the characteristic frequencies much smaller than those given by the bandgap is the so-called quasi-adiabatic Landau-Dykhne method~\cite{Delone2000}. Within this approach the ionization rate $W$ can be shown (see Methods) to be given as 
\begin{equation}
W\sim \mathrm{exp}\left(- 2 f(\gamma_K) \, \Delta/\omega\right),
\label{eq:W_main}
\end{equation}
 where $f(x)=\mathrm{arsinh}(x)
- (\sinh(2\mathrm{arsinh}(x))-2\mathrm{arsinh}(x))/(8x^2)$, and $\gamma_K$ is the so-called Keldysh parameter of the problem whose numeric value depends on the strength and the frequency of the driving field as well as on the details of the band structure of the material in question~\cite{Keldysh1965}. In the specific case of lead-bromide perovskites, density functional theory calculations provide $\gamma_K\simeq 0.85 \hbar \omega/(eaE_{AC})$, where $E_{AC}$ is the strength of the electric field induced by the laser, and $a\simeq 0.586$nm is the lattice constant MAPbBr$_3$ in cubic phase~\cite{Becker2018}.

The value of the Keldysh parameter determines the degree of adiabaticity of the problem, the two limiting cases being pure adiabatic tunneling ($\gamma_K \to 0$) and multiphoton absorption ($\gamma_K\to \infty$) as illustrated in Fig.~\ref{fig:fig1}A. {\color{black} Using the known $\mathbf{k}\cdot{\mathbf{p}} $ parameters of MAPbBr$_3$ in cubic phase~\cite{Becker2018}, we calculate in the limit $\gamma_K \to 0$: $W\sim \exp \left[   -{1.4\Delta}/{ (eaE_{AC})}    \right]$ (see Supporting Information), which  showcases the exponential dependence on the external electric field characteristic of the non-perturbative tunneling processes such as (static) Schwinger effect [cf.~Eq.~(\ref{eq:schwinger})]. The corresponding Schwinger field is $1.4\Delta/(ea)$, where the prefactor $1.4$ is specific for MAPbBr$_3$. In the multiphoton absorption regime~\cite{Nathan1985}, $W\sim (E_{AC})^{2\Delta/\omega}$. {\color{black}Note that pure adiabatic tunneling competes with the avalanche breakdown. Indeed, the condition
$\omega \gtrsim e E_{AC} /\sqrt{m \Delta}\simeq e a E_{AC}/\hbar$ is clearly violated in the limit $\gamma_K \to 0$. Therefore, in what follows, we focus on quasi-adiabatic tunneling with $\gamma_K\simeq 1$.}

For the experimental data presented in Fig.~\ref{fig:fig1}D, $E_{AC}$ is in between $0.24V/a$ and $0.36V/a$ (which corresponds to $E_{AC}\simeq 0.1\Delta/(ea)$). For that data $\hbar \omega\simeq 0.3$eV, so that $\gamma_K \simeq 1$. This puts the experiment in the quasi-adiabatic tunneling regime, i.e., the regime that extrapolates between Eq.~(\ref{eq:schwinger}) and multiphoton absorption.  As we argue below, the data in Fig.~\ref{fig:fig1}D is described quantitatively well using Eq.~(\ref{eq:W_main}) with the parameters of MAPbBr$_3$.
It is worth noting that we work with fields much smaller than the Schwinger field for our system. Nevertheless, we observe the PL signal because we are in a quasi-adiabatic regime, where the non-zero frequency of the laser reduces the field strength requirement.}

\subsection {Comparison to the experiment}
 In this section we confirm the tunneling nature of the ionization by deriving the functional dependence of the PL intensity, $I_{\mathrm{PL}}$, on the strength of the pumping mid-infrared (mid-IR) radiation. Naively, one might assume $I_{\mathrm{PL}} \propto W$, i.e., every single tunneling event results in a PL photon. However this is only true if:  1) there are no non-radiative recombination channels; 2) the photo-excited electron-hole pairs remain well-separated in space; 3) the density of charge carriers is low. Neither of these conditions are satisfied in our experiment. First of all, despite the excellent photoelectronic properties of LHPs the non-radiative recombination processes in them cannot be ignored altogether. Similarly, the long diffusion lengths of photocarriers in LHPs imply strong overlap between electron-hole pairs, which gives rise to a bi-molecular character of the recombination processes~\cite{Richter2016,Stranks2017}, {\color{black} see also Supporting Information for additional data in favor of a bi-molecular character.}
 
To quantitatively describe the photoluminescence in perovskites, one needs to know the population dynamics of photoexcitations $n(t)$ from which one can calculate the PL intensity based upon the bi-molecular recombination in lead halide perovskites, $I_{PL}\sim \int_{0}^t n(t)^2\mathrm{d}t$. The general equation governing the population of charge carriers $n(t)$ can be written as:
\begin{equation}
-d n(t)/dt =A_1 n + A_2 n^2 + A_3 n^3 + ....
\label{eq:kinetic}
\end{equation}

\noindent Here, the coefficients $\{A_i\}$ describe the rates of different decay channels of $n(t)$ such as: defect-assisted ($A_1$), bi-molecular (both radiative- and non-radiative; $A_2$), and Auger-type ($A_3$) recombination processes followed by higher-order processes, which should, in principle, be taken into account for sufficiently high densities of charge carriers, see Fig.~\ref{fig:fig1}E.

In the limit of weak pumping fields, $n(t)$ remains small at all times, implying that $n(t) = n(0) \exp\left(-A_1 t\right)$ and $I_{\mathrm{PL}} \sim n(0)^2$ where the initial population $n(0)$ is proportional to the tunneling ionization rate $W$. This result is accurate as long as $n_0\ll A_1/A_2$. Note that $A_1/A_2\simeq 3.5\times 10^{16}$cm$^{-3}$~\cite{Richter2016}.
As the pumping intensity grows, the omission of higher-order terms in Eq.~(\ref{eq:kinetic}) is no longer justified. For example, the inclusion of both mono- and bi-molecular channels leads to $I_{PL}\sim \alpha W-\mathrm{ln}(1+\alpha W)$, where $\alpha$ is a fitting parameter (see Methods). This expression is accurate for $E_{AC}\lesssim 0.3 V/a$,  providing an estimate for a number of excited particles in the system $10^{16}-10^{17}$cm$^{-3}$ at this field. For higher excitation densities, the channels beyond bi-molecular recombination have to be included. 

In Fig.~\ref{fig:fig1}D we show the results of the fit limited to second (bi-molecular) processes (yellow curve) as well a phenomenological fit including higher-order channels (red curve) line. The phenomenological fit is motivated by our analytical results. It has the form 
$I_{PL}\sim \alpha \tilde W-\mathrm{ln}(1+\alpha \tilde W)$, where $\tilde W=W/(1+\beta  W)$ is a renormalized ionization rate that takes into account that for large densities only a fraction of charge carriers can recombine radiatively. Note that the phenomenological fit has now two parameters $\alpha$ and $\beta$. The overall prefactor that connects $I_{PL}$ to $n^2$ is beyond the Landau-Dykhne approach. {\color{black} It can be treated as another fit parameter, which in a log-plot simply shifts the data along the vertical direction. Since the exact relationship between the total number of PL photons and the number of photons detected is generally unknown, the prefactor can be entirely omitted by properly normalizing the data, see Fig.~\ref{fig:fig1}D.}

 {\color{black}
\section{Applications of tunneling ionization}
}

%%%%%%%%%%%%%%%%%%%%%%%%%%%%%%%%%%%
\begin{figure}[t]
\includegraphics[scale=0.5]{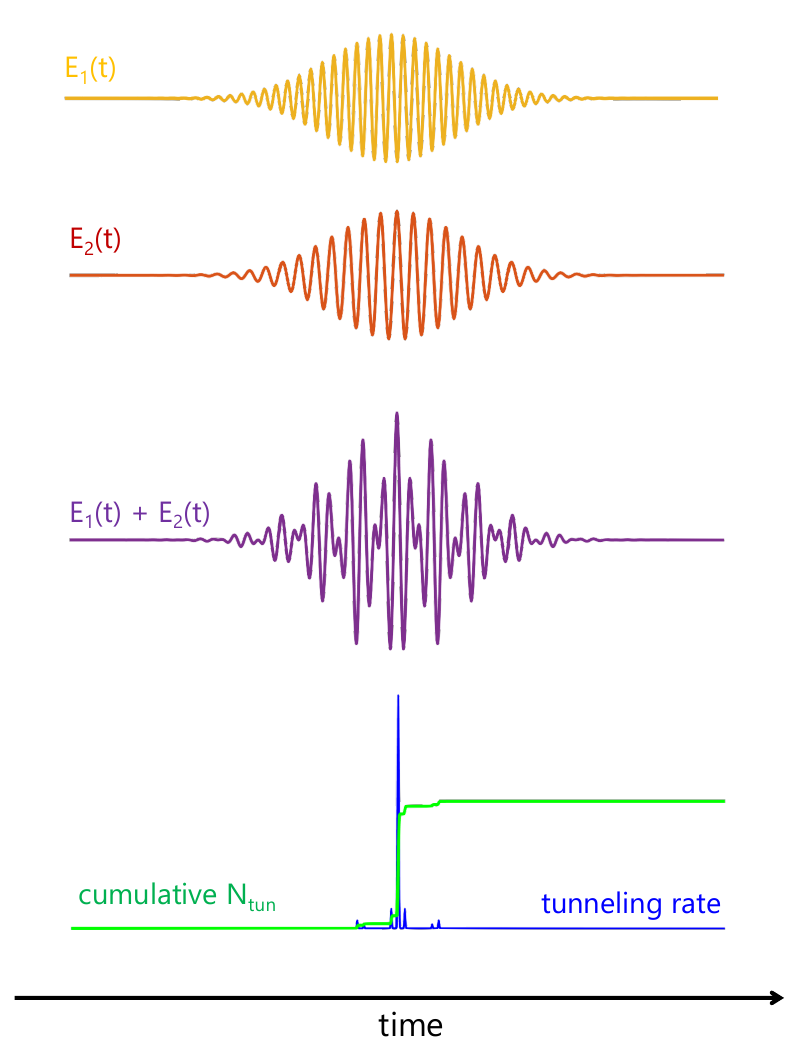}
\caption{
{\color{black} Sketch of tunneling under the influence of two slowly varying in time fields $E_1(t)$ and $E_2(t)$ (yellow and red curves at the top of the figure). Their sum $E_1(t)+E_2(t)$ is also presented (magenta). The tunneling rate (blue at the bottom of the figure) is exponentially enhanced at the maximum of the total field. Consequently the cumulatative number of transitions (green) is determined mainly by the value of this maximum.}
}
\label{fig:fig2_resub}
\end{figure}
%%%%%%%%%%%%%%%%%%%%%%%%%%%%%%%%%%%

{\color{black}The hallmark feature of quantum tunneling is its exponential sensitivity to external parameters, which has enabled applications defying the classical common sense. The most celebrated example is the scanning tunneling microprobes where exponential sensitivity of the tunneling current means that the overwhelming majority of it is flowing through the single atom that happens to stick out the most toward the sample ~\cite{Binnig1982, Chen2007}. Realization of this has led to unprecedented sub-atom-scale spatial resolutions. Other celebrated applications of tunneling-related process include the tunnel diode~\cite{Turner1976} with its unique I-V characteristics and high-harmonic generation in atomic optics where deeply sub-threshold laser light promotes electrons in an atom to unoccupied states~\cite{Yue2022}, thus initializing the process of extreme frequency conversion of radiation. 

Likewise, besides being a fascinating example of quantum tunneling dynamics, our findings are also of practical importance. The inherent non-perturbative character of tunneling in our case leads to a possibility to amplify the effect of weak fields by driving the system with another laser field.} The basic intuition here can be inferred already from the expression in Eq.~(\ref{eq:schwinger}), which suggests that the overwhelming majority of tunneling events occurs near the moment of time $t_0$ when the total field $E_{tot}(t)$ reaches its maximum $E_{tot}(t_0) = E_{max}$.  For instance, in the case of tunneling under the simultaneous influence of two slowly changing electric fields $E_1(t)= E_1 \cos(\omega_1 t),$ and $E_2(t)=E_2 \cos(\omega_2 t + \phi)$, the total number of tunneling events according to this logic is expected to be $N_{tun} = \int W(t) dt \propto \exp \left( -E_s/(|E_1|+|E_2|) \right)$ {\color{black}(see an illustration in Fig.~\ref{fig:fig2_resub})}. While this simple example is valid only for $\omega_1, \omega_2 \to 0$ it in fact illustrates the more general phenomenon of exponential cooperative enhancement of tunneling yield. A more rigorous treatment of this effect can be provided within the quasi-adiabatic formalism~\cite{Perelomov1967}. Below, we illustrate that the exponential cooperative enhancement can indeed be used a valuable resource. 

\subsection{{\color{black}Detection of polarization-dependent enhancement of tunneling in LHP}}

The distinctive exponential sensitivity of tunneling ionization rate on the peak value of the {\it net} applied field can be used as a resource to detect weak local fields in a material. This can be best illustrated in the quasi-static regime describe above. Using the Landau-Dykhne approach it can also be demonstrated in a general case that even a weak static frozen-in field $E_1$ ($\omega_1=0$) can strongly enhance the tunneling rate $W$ due to the additional driving field $E_2(t)$ (see Supporting Information). The fact that local fields can enhance PL yield is consistent with a qualitative observation that
we see stronger PL in the vicinity of structural defects (edges, surfaces) of LHPs where internal electric fields, $E_1$, are expected to appear~\cite{Niesner2016}.

As an application of this sensitivity of PL to $E_1$ we propose to use it to study the highly debated frozen-in electric fields $E_{DC}$ in LHPs~\cite{Ambrosio2022}, which lead to a broken inversion symmetry even for nominally cubic lead-halide perovskites. It is worth noting here that the method merits certain advantages as compared to some of the conventional techniques for detecting local breaking of inversion symmetry such as second-harmonic generation or angle-resolved photoemission. First of all, our method is not confined to the vicinity of the sample surface since the used electric fields have sub-gap frequencies and hence can penetrate the sample. Second, the peculiarity of tunneling is that it is sensitive to the {\it absolute value} of field amplitudes, cf. Eq.~(\ref{eq:schwinger}). Therefore there is a finite net effect due to local inversion breaking even if the local fields spatially average to zero.
Finally, another implication of the field cooperation is that PL should strongly depend on the polarization of the external field.

\begin{figure}[t]
\includegraphics[scale=0.4]{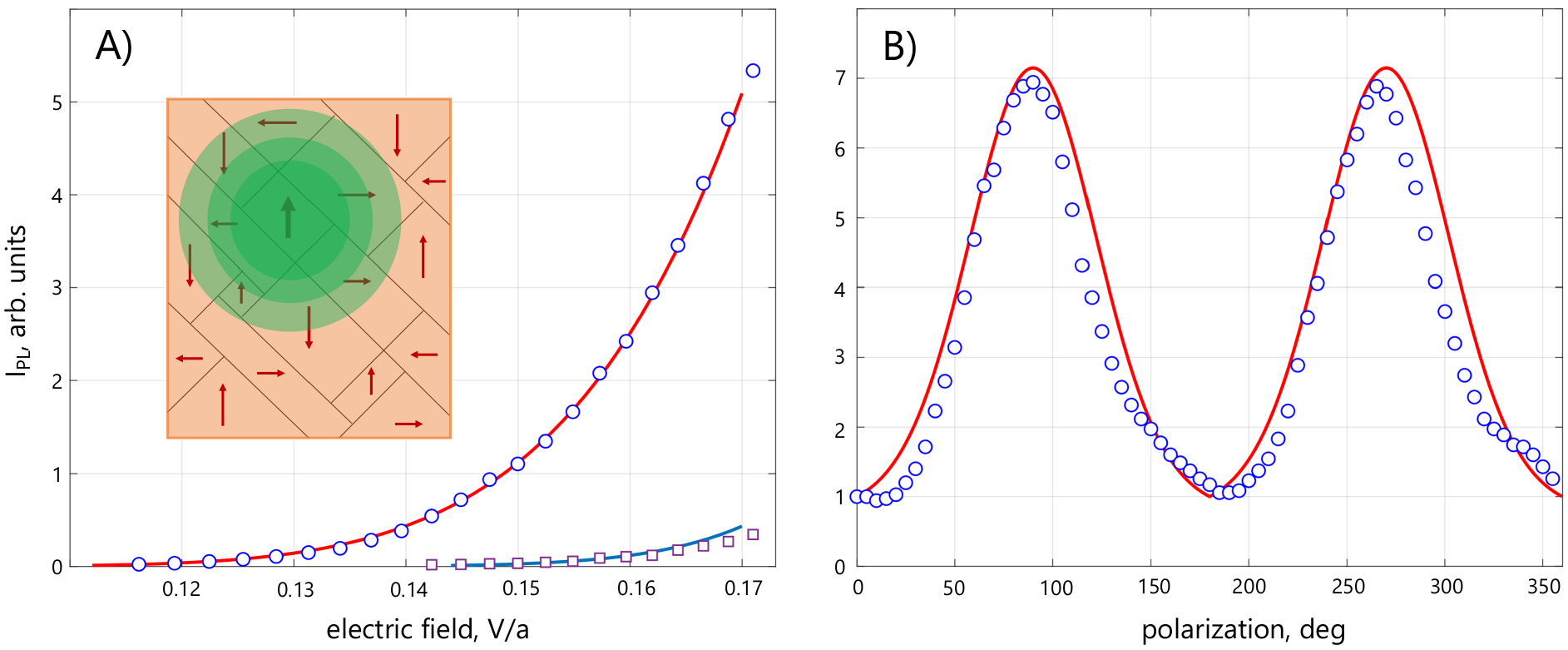}
\caption{A) PL as a function of the applied electric field for two orthogonal polarization orientations aligned with the crystal axes of MAPbBr$_3$ (markers). The corresponding fits according to our theoretical model to the data are presented as solid curves. The inset shows a cartoon of MAPbBr$_3$ with hypothetical ferroelectric domains overlaid with a schematic representation of the irradiating laser spot whose diameter is $d\approx 200\mu$m ($1/e^2$); B) Polarization scan taken at a fixed value of the external electric field  $E_{AC}\simeq 0.180 V/a$.}
\label{fig:fig3}
\end{figure}

 For experimental validation, we measure PL from a single-crystal MAPbBr$_3$ as a function of intensity of the driving mid-infrared $\lambda = 4\mu$m radiation. The resulting dependence plotted in Fig.~\ref{fig:fig3}A indicates that the tunneling ionization efficiency depends strongly on the polarization of the pumping field. 
 To elucidate the nature of this polarization dependence, we scan PL as a function of polarization at a fixed pump intensity as shown in Fig.~\ref{fig:fig3}B. 

The two-fold symmetric polarization dependence in Fig.~\ref{fig:fig3}B might appear surprising since this measurement was performed at room temperature where MAPbBr$_3$ is expected to be in the cubic phase, i.e., four-fold symmetric (e.g., consider the two-photon absorption experiments like in Ref.~\cite{Li2020Nano}). The data in Fig.~\ref{fig:fig3}B therefore indicates an unexpected lowering of symmetry in MAPbBr$_3$ in the nominally cubic phase. In the most straightforward explanation this could be attributed to an extrinsic factor such as residual stress in the crystal structure that would lift the four-fold symmetry of the electronic band structure. However it can be demonstrated that, that in this case in order to reproduce the magnitude of the effect in Fig.~\ref{fig:fig3} the intrinsic stains in the system have to reach relatively large ($\simeq 0.5$\%) levels~\cite{Chen2020}, which are not expected in a single-crystal LHP~\cite{Li2024} (see Supporting Information for more details).

%%%%%%%%%%%%%%%%%%%%%%%%%%%%%%%%%%%
\begin{figure*}[t]
\includegraphics[scale=0.5]{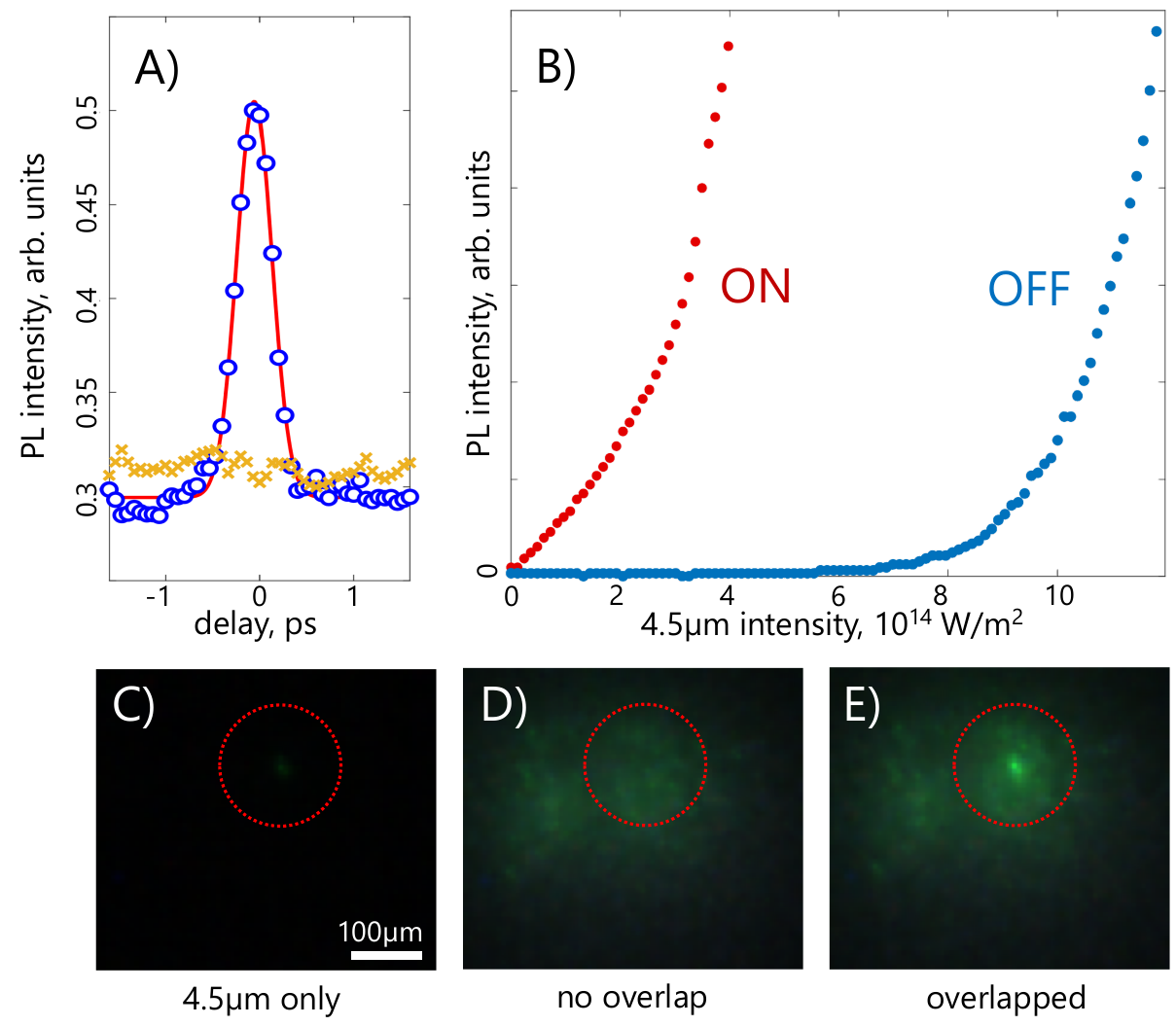}
\caption{
{\color{black} A) Cumulative PL from a single-crystal sample MAPbBr$_3$  as a function of the time delay between 1$\mu$m and 4$\mu$m pulses with parallel- (blue circles) and orthogonal (yellow crosses) polarizations; solid red line is a Gaussian fit to the curve used as the guide to an eye; B) Differential PL from a single-crystal sample MAPbBr$_3$ as a function of mid-infrared intensity (4.5$\mu$m) biased with $1\mu$m at $I_{\mathrm{bias}}=5\times10^{13}$W/m$^2$ (mid- and near-infrared pulses overlap in time; red points) as compared to PL induced by the same mid-infrared beam in the absence of AC-biasing (mid- and near-infrared pulses do not overlap in time; black points). {\color{black} Panels C)-E) demonstrate spatial profiles of PL under AC-biasing in an abraded sample of MAPbBr$_3$. Namely, panel C) shows PL produced by the pulse $4.5\mu$m alone; panels D) and E) present spatial profiles of cumulative PL with $1\mu$m and $4.5\mu$m pulses not overlapped and overlapped in time, respectively}. The red circle in C)-E) marks the position of the $4.5\mu$m light beam.}
}
\label{fig:fig4}
\end{figure*}
%%%%%%%%%%%%%%%%%%%%%%%%%%%%%%%%%%%

In an alternative scenario, the observed anisotropy can be attributed to the local frozen-in electric fields, that were conjectured to be present in lead-halide perovskites~\cite{Niesner2016,Rakita2017}. Indeed the exponential sensitivity of PL on applied fields implies that one can obtain the anisotropy observed in Fig.~\ref{fig:fig3} already with moderate static fields. In Figs.~\ref{fig:fig3}A and B we show that all of the experimental data can be reproduced within the two-field Landau-Dykhne formalism presented in the Supporting Information. Based on the fitting, we estimate the numeric value for the internal static fields in room-temperature single-crystal MAPbBr$_3$ to be of the order of $E_{DC} \sim 0.1$V/nm. Within this interpretation of the PL anisotropy, the PL gets maximum enhancement when the polarization of the driving optical field coincides with the direction of $E_{DC}$. That this direction happens to be aligned with direction of Pb-Br bonds ([001] of the cubic lattice) is consistent with previous observations of possible ferroelectricity in MAPbI$_3$~\cite{Rhm2019, Leonhard2019}. Note that Fig.~\ref{fig:fig1}D is produced using electric fields that are stronger than those presented in Fig.~\ref{fig:fig3}A. Further, we rotated the polarization of the external electric field to have minimal PL intensity. This allowed us to minimize the effect of the frozen-in electric fields in the data demonstrated in Fig.~\ref{fig:fig1}D. 

Importantly, we also observe that the two-fold symmetric pattern of PL is subject to the spatial extent of the focal spot of the driving laser. Namely, we observe that the two-fold symmetry in Fig.~\ref{fig:fig3} is only present for sufficiently focused beams. For larger illumination spots, the pattern turns out to be four-fold symmetric (see Supporting Information). This can be naturally explained if more than one ferroelectric domain with different field directions participate in the fluorescence process. This interpretation agrees well with the previous estimations of the size of possible ferroelectric domains of several microns in MAPbI$_3$~\cite{Rothmann2017, Rakita2017}.

\subsection{Cooperation of dynamic fields and AC-biasing} 
\label{subsec:cooperation}

The cooperation effect illustrated in Fig.~\ref{fig:fig2_resub} implies that the total yield of tunneling ionization across the band gap driven by an AC-field can be significantly enhanced in the presence of an additional ``boosting'' AC-field, which does not have to be in any particular phase or frequency relationship to the first field. In order to demonstrate this ``AC-biasing'' phenomenon, we measure the PL from the MAPbBr$_3$ sample irradiated by a 4.5$\mu$m beam while it is being ``AC-biased'' by an additional 1$\mu$m beam, {\color{black}see Experimental section for details}. 

{\color{black}In Fig.~\ref{fig:fig4}A we show the cumulative PL, $\int_{-\infty}^\infty \mathrm{d}t I_{\mathrm{PL}}(t,\tau)$, coming from the sample as a function of the delay time ($\tau$) between the two pulses in cross- and parallel polarization geometries (yellow x- and blue o-markers, respectively)}. As can be clearly seen, the PL signal is enhanced only when the two pulses of the same polarization overlap in time, i.e., when the amplitude of the sum field is maximized in full qualitative agreement with the concept of the dynamical assistance in quantum tunneling introduced in the previous sections. 

As a demonstration of the practical use of this effect, we present in Fig.~\ref{fig:fig4}B the effect of AC-biasing on PL produced by a mid-infrared beam ($4.5\mu$m) in MAPbBr$3$ biased with $1\mu$m radiation at $I_{\mathrm{bias}}=5\times10^{13}$W/m$^2$, {\color{black}see Supporting Information for additional data}. Here, we show the differential photoluminescence for two beams  $\Delta PL(I) = PL(I_1, I_{\mathrm{bias}}) -  PL(I_1, 0) - PL(0, I_{\mathrm{bias}})$, where $PL(I_1, I_{\mathrm{bias}})$ stands for the total PL emitted by the sample when irradiated by mid- and near-infrared pulses overlapping in time. As can be seen the differential sensitivity of PL can be made linear in the intensity of the mid-IR which opens broad possibilities for efficient mid-infrared sensors based upon lead-halide perovskites. 

{\color{black} There is a caveat to consider when designing such a sensor. All the results presented thus far correspond to measurements on a single crystal, away from any structural features like crystal edges. This approach was taken because, as demonstrated above, the efficiency of tunneling ionization is highly sensitive to the local properties of the sample (e.g., stress, static fields). For instance, we observe that photoluminescence (PL) becomes visibly enhanced when the irradiated region is near structural defects such as surface inhomogeneities or sample edges/cracks. This is problematic for potential infrared imaging applications, as structure-induced variations in PL may obscure the actual spatial distribution of the mid-infrared wavefront we aim to probe. 

Sample imperfections are unavoidable in practice. Therefore, not being able to ensure sample homogeneity by removing the imperfections one can take the other extreme and instead homogenize the defect distribution, thus obtaining a quasi-uniform sample suitable for infrared-imaging purposes. To demonstrate this approach in practice, we deal with abraded powder samples (see Supporting Information). Although, the intrinsic in-homogeneity of these samples does not allow for a quantitative description, they can be directly used for sensing  as we demonstrate by overlapping the two beams (mid- and near-infrared) on a screen covered by the abraded sample. }
As can be seen in Fig.~\ref{fig:fig4}C-E, there is clear enhancement in local PL from the region of spatial overlap. Curiously, due to the nonlinear sensitivity of PL on the pumping beam intensity, the apparent size of the overlap spot is significantly smaller (11 $\mu$m) than the actual spot size of the mid-infrared pulse (160 $\mu$m), which can be employed for super-resolution microscopy.

\section{Summary}

To summarize, we have demonstrated tunneling ionization in a Dirac semiconductor, quantitatively described it with the quasi-adiabatic Landau-Dykhne approach, and interpreted it in terms of {\color{black}analogue} dynamical Schwinger effect. {\color{black} This became possible  due to unique photoelectric properties of lead halide perovskites, particularly due to their very high quantum PL efficiency, which enables a detection of tunneling-ionized carriers through PL long before impact ionization becomes relevant. Furthermore, a considerable energy gap of MAPbBr$_3$ allowed us to easily rule out all upconversion mechanisms based upon low energy scales of the material, for example, a phonon-assisted upconversion~\cite{Ye2021,Dai2023}. Indeed, one would require of the order of $50$ phonons to account for a difference between the frequencies of the driving field and the PL signal.

Utilizing the exponential sensitivity of tunneling ionization to driving fields, we measured the polarization dependence of the PL signal and interpreted the results by invoking local frozen-in fields in a nominally cubic MAPbBr$_3$ single crystal at room temperature. These results provide new insights into the ongoing debate on polar order in lead-halide perovskites.} Finally, we have investigated the cooperation between two time-dependent fields simultaneously driving tunneling ionization and demonstrated that this cooperation can act as an AC analogue of biasing for an optical frequency {\color{black}(non-parametric)} upconversion. These findings pave the way for a mid-infrared light detection with lead-halide perovskites.

%\appendix 

\section{Experimental section/Methods}

\subsection{Experimental setup}
\label{app:a}

{\color{black}
\noindent The experimental setup, see Fig.~\ref{fig:setup}, consisted of an amplified femtosecond laser system (Light Conversion PHAROS) coupled to an optical parametric amplifier (OPA, Light Conversion ORPHEUS). The laser produces a train of pulses centered at 1028nm with a repetition rate of 3 kHz, pulse duration of 300 fs and a pulse energy of 2 mJ. A small fraction (5\%) of the main beam was split off and used as a NIR probe while the main part pumped the OPA producing a  MIR pump beam. Pump and probe pulses were spatially and temporally overlapped inside the LHPs. The resulting PL was sampled with an amplified silicon photodetector (PDA-100A2, Thorlabs). A short pass filter (Thorlabs) was used in order to cut off the remaining 1030 nm light. 

\begin{figure}
  \includegraphics[width=0.7\linewidth]{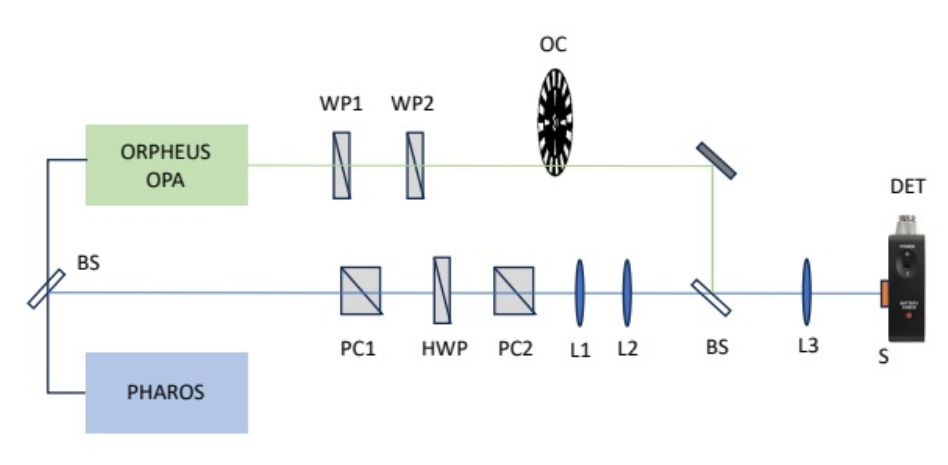}
  \caption{{\color{black}Experimental Setup, BS - beamsplitter, WP - wiregrid polarizer, OC - optical chopper, PC - cube polarizers, HWP - half-wave plate, L - lens, S - sample.}}
  \label{fig:setup}
\end{figure} 

\noindent \textit{The PL spectra} were taken in reflection geometry by pumping the sample with respective wavelengths at an intensity of $1\times10^{11}$\,W/cm$^2$ and sampling the PL with a fiber coupled spectrometer (OceanOptics FLAME-T).

\noindent {\it The beam diameters} were characterized for both the MIR and NIR beams. For the former we employed the knife edge technique~\cite{Khosrofian1983}, while for the latter a direct imaging by a CMOS camera has been applied. In Fig.~\ref{fig:modes1}, we illustrate the knife-edge data for the pump wavelength of 4 $\mu$m. It corresponds to the measurement reported in Fig.~\ref{fig:fig1} of the dependence of PL intensity on the magnitude of the external laser field.  

Similarly in the two-color experiment (see Fig.~\ref{fig:fig4}) that was performed by mixing the 1 $\mu$m and 4 $\mu$m fields, we characterized the spot size of the latter by the knife-edge technique while for the former we used a direct imaging by a CMOS camera. We extracted a beam radius of 113 $\mu$m for the MIR beam and a transverse radius of 83 $\mu$m $\times$ 121 $\mu$m for the elliptical NIR beam, see Fig.~\ref{fig:modes2}. Note that we used spatial filtering in the MIR arm and hence the beam was almost perfectly radially symmetric in this case.

\noindent {\it Polarization resolved scans} were performed by first converting the linearly polarized MIR radiation from the OPA into a circularly polarized state by means of a tunable quarter-wave plate (ALPHALAS) and a subsequent rotation of the polarization plane by a wiregrid polarizer (THORLABS).  

\noindent {\it In the two-color experiment} a pair of half-wave plate and Glan-Taylor polarizer (GT10, Thorlabs) and a pair of wiregrid polarizers were used to continuously tune the incident power in the 1030 nm and 4500 nm arms respectively. For imaging the MAPbBr$_3$ sample was placed in a defocused 1030 nm beam while being kept in the focus of the 4500 nm pump beam. The images were then taken using a CMOS camera (Point Grey Research PGR-CM3-U3-50S5M-CS).

}

\begin{figure}[t]
  \includegraphics[width=0.5\linewidth]{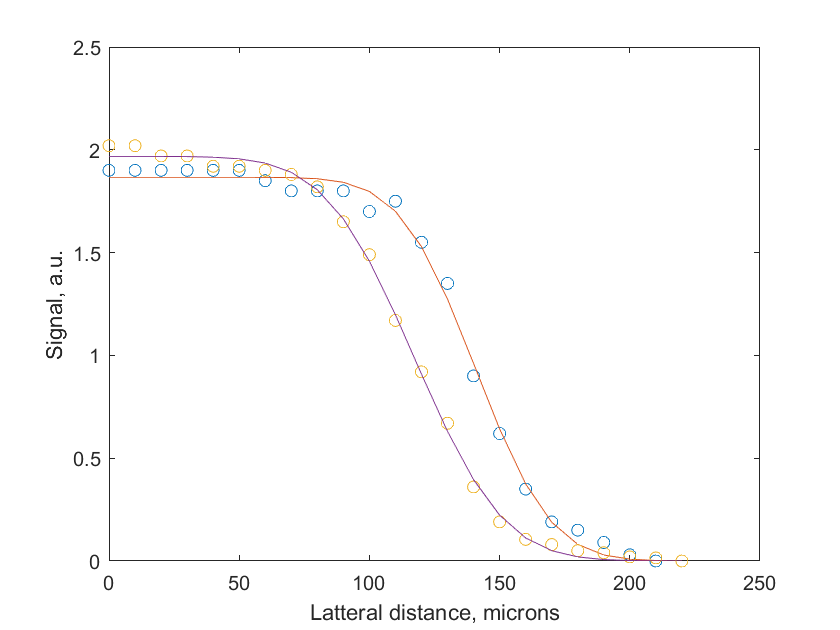}
  \caption{{\color{black}The 4 $\mu$m pump beam spot size as characterized by the knife-edge technique. 
   The blue circles correspond to the scan in the horizontal plane while the yellow circles mark the scan in the vertical plane. From the latter we extracted the beam radii of 44 $\mu$m and 53 $\mu$m, respectively. These values were subsequently used to evaluate the peak intensity and analyze the data presented in Fig.~\ref{fig:fig1}.} }
  \label{fig:modes1}
\end{figure}

\subsection{Calculation of photoluminescence}
\label{app:b}

 \begin{figure}[t]
  \includegraphics[width=0.7\linewidth]{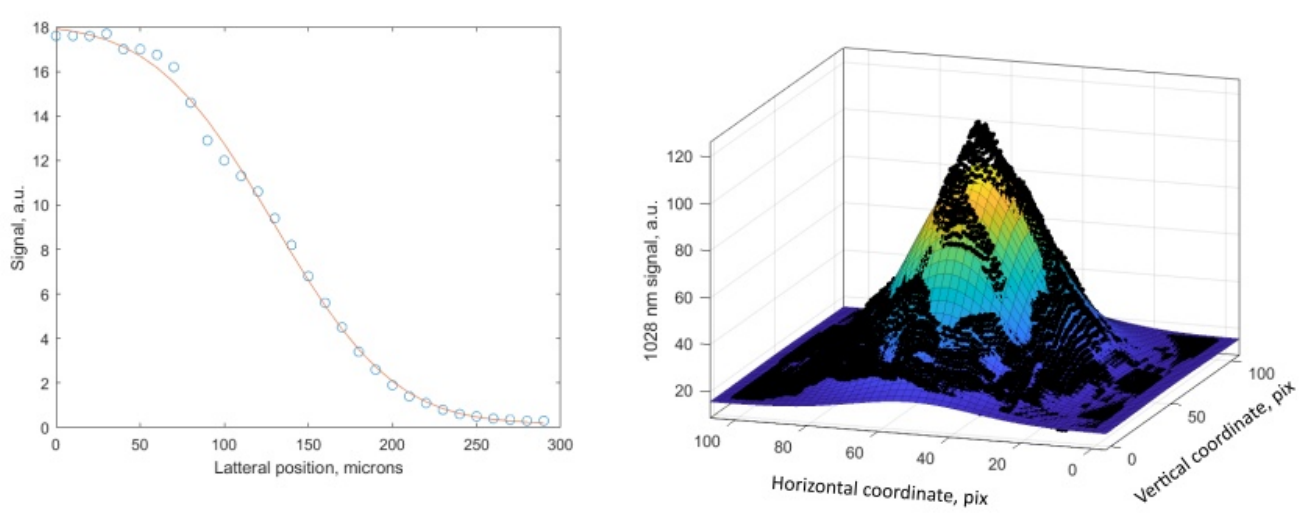}
  \caption{{\color{black} The 4 $\mu$m pump beam spot size as characterized by the knife-edge technique (left) and the 1 $\mu$m beam spot as imaged by a CMOS camera (right), see the text for more details.}}
  \label{fig:modes2}
\end{figure}

\noindent To understand the parametric dependence of the PL, we use the Landau-Dykhne adiabatic approximation where the probability of transition from the initial state $i$ to the final state $f$ is given by the expression($\hbar=1$)~\cite{Delone2000}
\begin{equation}
W_{fi}\sim \mathrm{exp}\left(-2\mathrm{Im}\int_0^T (\mathcal{E}_{f}(t)-\mathcal{E}_{i}(t))\mathrm{d}t\right),
\label{eq:Wmethods}
\end{equation}
where $\mathcal{E}(t)$ is the instanteneous energy of the time-dependent Hamiltonian, $H$; $T$ is the (complex) instance of time when $\mathcal{E}_{f}(t)=\mathcal{E}_{i}(t)$. To calculate the energies, we
 use the following Hamiltonian~\cite{Jin2012,Becker2018}
\begin{equation}
H=\frac{1}{2}\left(\Delta+t_3 \frac{(\mathbf{k} a)^2}{2}\right)\tau_3\otimes \sigma_0 + 2t a \tau_2 \otimes \sum_{l=1}^3 \sigma_l k_l,
\label{eq:methos_Ham}
\end{equation}
which describes the band structure in the vicinity of the bandgap, i.e., $k\to 0$.  
Here, $\Delta\simeq 2.4$eV is the energy gap between the conduction and valence bands; $t\simeq 0.6$eV and $t_3\simeq 0.9$eV are the hopping integrals; $a\simeq 0.586$nm is the lattice spacing; 
$\tau_i$ and $\sigma_i$ are the Pauli matrices acting on the orbital and quasispin degrees of freedom, respectively. Further details of the used notation can be found in Ref.~\cite{Volosniev2023}. Note that we do not include the spin-electric term~\cite{Volosniev2023a} in our calculations, because its contribution will be subleading for $\omega\to 0$ {\color{black}(see Supporting Information)}.  
External fields enter Eq.~(\ref{eq:methos_Ham}) via the minimal coupling substitution $\mathbf{k}\to \mathbf{k}-e\mathbf{A}$; we shall assume weak fields so that $e a\mathbf{A}\to 0$.
We are interested in the regime that is exponentially sensitive to the parameters. Therefore, 
we shall consider $\mathbf{k}=0$, which determines the most probable excitation process. With these approximations, the Hamiltonian of interest reads
\begin{equation}
H=\frac{1}{2}\left(\Delta+\frac{t_3 (e a\mathbf{A})^2}{2}\right)\tau_3\otimes \sigma_0 + 2 e a t \tau_2 \otimes \sum_{l=1}^3\sigma_l A_l.
\end{equation}
Its energies are 
\begin{equation}
\mathcal{E}_{i}=-\sqrt{\frac{1}{4}\left(\Delta+\frac{t_3 (e a\mathbf{A})^2}{2}\right)^2 + 4 t^2 (e a \mathbf{A})^2},
\label{eq:energies}
\end{equation}
and $\mathcal{E}_{f}=-\mathcal{E}_{i}$. Note that the energy states are double degenerate. 
Therefore, in general, we should define a conserved quantity -- quasi-spin, and consider separate quasi-spins in parallel. However, as Eq.~(\ref{eq:Wmethods}) is independent of this quantum number, we will not take this degeneracy into account.
To calculate the value of $\tau$ that appears in Eq.~(\ref{eq:Wmethods}), we solve the equation $\mathcal{E}_{i}=0$, which leads to
$(e a\mathbf{A})^2 \simeq -\Delta^2/(\Delta t_3+16 t^2)$. This expression together with Eqs.~(\ref{eq:Wmethods}) and~(\ref{eq:energies}) 
provide the basis for our calculations of PL in the main text.

As an example, let us calculate the PL intensity for a monocromatic light beam.
We assume that the vector potential has the form 
$\mathbf{A}=-\mathbf{E}_{AC}\sin(\omega t)/\omega$,
where $\mathbf{E}_{AC}$ is a constant vector that defines the strength of the electric fields (recall that $\mathbf{E}=-\partial \mathbf{A}/\partial t$ in SI units).
It is clear that to satisfy $\mathcal{E}_{f}(t)=\mathcal{E}_{i}(t)$, $t$ should be imaginary, i.e., $t=i\tau$, where
\begin{equation}
 \tau= \frac{1}{\omega}\mathrm{arsinh}\left(\frac{\omega}{eaE_{AC}} \sqrt{\frac{\Delta^2}{\Delta t_3+16 t^2}}\right). 
\end{equation}
Now, we have all ingredients to calculate $W_{fi}$:
\begin{equation}
W_{fi}\sim \mathrm{exp}\left(-2\mathrm{Im}\int_0^{i\tau} \left[\Delta+\frac{\Delta t_3 + 16 t^2}{2\Delta}\left(ea \mathbf{A}\right)^2\right]\mathrm{dt}\right),
\nonumber
\end{equation}
which leads to Eq.~(\ref{eq:W_main}) of the main text with the Keldysh parameter for our problem
\begin{equation}
\gamma_K=\frac{\omega}{eaE_{AC}} \sqrt{\frac{\Delta^2}{\Delta t_3+16 t^2}}.
\end{equation}

\subsection{PL intensity at strong fields}
\label{app:c}

\noindent Let us first consider the situation when $n_0$ is of the order of $A_1/A_2$, but still much smaller than $A_2/A_3\simeq 10^{18}$cm$^{-3}$~\cite{Richter2016}. In this case 
$\mathrm{d} n/\mathrm{d}t=-A_1 n-A_2 n^2$, which leads to $n=C e^{-A_1 t}/(1-A_2 C e^{-A_1t}/A_1)$, where $C=n_0/(1+A_2 n_0/A_1)$. The corresponding PL intensity reads
\begin{equation}
I_{PL}\sim \left(n_0+\frac{A_1}{A_2}\mathrm{ln}\left[\frac{A_1}{A_1+A_2n_0}\right]\right).
\end{equation}
Noticing that $W_{fi}\sim n_0$, we recover the result presented in the main text: $I_{PL}\sim \alpha W_{fi}-\mathrm{ln}(1+\alpha W_{fi})$,
where $\alpha$ is a fitting parameter. Inclusion of the processes with $i>2$ in the rate equation will lead to more fitting parameters, and, correspondingly, to a better agreement between the theory and the data. From a physical point of view higher-order processes suppress two-body losses, and effectively renormalize the initial density for the radiative recombination. Phenomenologically, this can be easily included using the following expression $I_{PL}\sim \frac{\alpha W_{fi}}{1+\beta  W_{fi}}-\mathrm{ln}\left(1+\frac{\alpha W_{fi}}{1+\beta W_{fi}}\right)$. This expression with the fitting parameters $\alpha$ and $\beta$ reproduces our data well everywhere.

%\clearpage

\textbf {Supporting Information} Supporting information is available from the publisher or from the author. 

{\color{black}Sample preparation; Numerical calculation of tunneling ionization; Ionization in the presence of a frozen-in electric field; Additional information in support of the bi-molecular-recombination origin of photoluminescence;
Additional information for the discussion on frozen-in electric fields; Hysteresis of photoluminescence intensity; Abraded samples of MAPbBr$_3$; Additional data for a two-color experiment; Photoluminescence from CsPbBr$_3$. Supporting information contains additional references~\cite{Abiedh2021,Kumar2025}.}

\textbf {Acknowledgements} A. G. V. thanks Peter Balling for useful discussions.

\textbf {Contribution} D.L. and A. G. V. contributed equally.

%\bibliographystyle{plainnat}
%\bibliography{ref}

\newpage
\clearpage

\textbf {Supporting Information}

\vspace{3em}

\noindent \textbf{Sample preparation}

\noindent \textit{Chemicals} CH$_3$NH$_3$Br (>99.99\%) was purchased from GreatCell Solar Ltd. (formerly Dyesol) and used as received. PbBr$_2$ (98\%), CsBr (99.9\% trace metals basis), DMF (anhydrous, 99.8\%), and DMSO (anhydrous, 99.9\%) were purchased from Sigma Aldrich and used as received.

\noindent \textit{Synthesis} of MAPbBr$_3$ perovskite single crystals. A 1.5 M solution of CH$_3$NH$_3$Br/PbBr$_2$ in DMF was prepared, filtered through a 0.45 {\textmu}m-pore-size PTFE filter; and the vial containing 0.5-1 ml of the solution was placed on a hot plate at 30 $^\circ$C. Then the solution was gradually heated to 60 $^\circ$C and kept at this temperature until the formation of MAPbBr$_3$ crystals. The crystals can be grown into larger sizes by elevating the temperature further. Finally, the crystals were collected and cleaned using a Kimwipe paper. 
CsPbBr$_3$ perovskite single crystals were synthesized following a procedure reported before~\cite{Saidaminov2017}.

\vspace{1em}
{\color{black} 
\noindent \textbf{Numerical calculation of tunneling ionization} 
\vspace{1em}

In calculation of Eq. (2) of the main text, we used a number of approximation, see the Methods of the main text.
In particular, we used a quasi-adiabatic approximation and disregarded the spin-electric term~\cite{Volosniev2023a}, also known as Pauli coupling~\cite{Kumar2025}.
Although these approximations appear natural, here we further validate them by comparing our analytical expression 
to a numerical solution. To this end, we solve numerically the Schr{\"o}dinger equation 
\begin{align}
i\hbar\frac{\partial \alpha}{\partial t}=\left(\frac{1}{2}\Delta+t_3 \frac{(eaA)^2}{4}\right)\alpha+\left(\mu E + 2at e Ai\right)\beta,\\
i\hbar\frac{\partial \beta}{\partial t}=-\left(\frac{1}{2}\Delta+t_3 \frac{(eaA)^2}{4}\right)\beta+\left(\mu E - 2at e Ai\right)\alpha,
\label{eq:system_of_eqs}
\end{align}
where $\alpha$ abd $\beta$ represent the amplitudes of the wave function for the conduction and valence bands, respectively. Without loss of generality, we assumed that $\vec A=(0,0,A)$.
We aim to find $|\alpha(t\to\infty)|^2$ subject to the initial conditions $\alpha(t\to-\infty)=0$ and $\beta(t\to-\infty)=1$.
To streamline the analysis, we make a number of assumptions. 
First, we use $A(t)=E te^{-\omega^2 t^2}$. This vector potential will lead to a unique pole for the Landau-Dykhne approach,
facilitating a direct comparison between the methods. Second, we will set $t_3=0$ for simplicity.
Otherwise, we use the parameters of the main text.

Figure~\ref{eq:system_of_eqs} presents a comparison between the numerically exact solution of Eq.~(\ref{eq:system_of_eqs}) and the Landau-Dykhne approximation for different values of $\omega$ (recall that $\omega/\Delta$ governs the degree to which the process is adiabatic). Following the discussion in the main text, we set $\mu=0$ in the numerical calculations.
We see that for $\hbar\omega\simeq 0.21$eV the Landau-Dykhne approximation is fully justified for the considered values of $E$. There is virtually no difference between numerical and analytical results. The case with $\hbar\omega\simeq 0.31$eV, which corresponds to our experiment, is also within the reach of the quasi-adiabatic approximation (a slight discrepancy between analytical and numerical results is unlikely to be resolved with the present experimental accuracy).

\begin{figure}
 \subfigure[]{\includegraphics[width=0.45\textwidth]{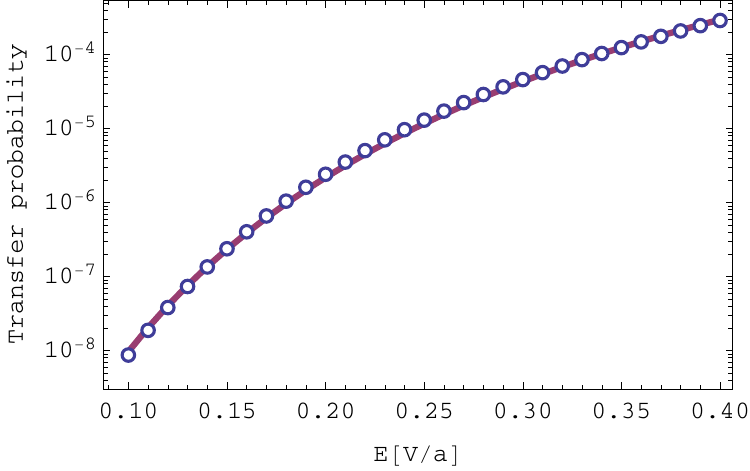}} 
  \subfigure[]{\includegraphics[width=0.45\textwidth]{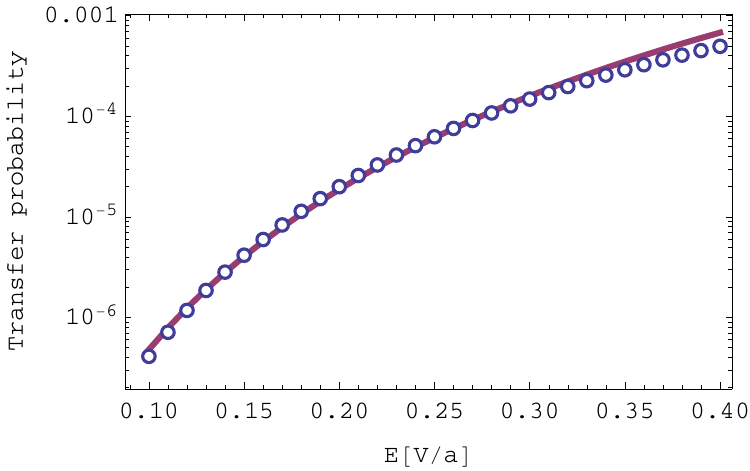}} 
  % \subfigure[]{\includegraphics[width=0.3\textwidth]{omega_0.35.pdf}} 
%\begin{subfigure}{0.33\textwidth}
%\includegraphics[width=0.9\linewidth]{omega_0.21.pdf} 
%\end{subfigure}
%\begin{subfigure}{0.33\textwidth}
%\includegraphics[width=0.9\linewidth]{omega_0.31.pdf}
%\end{subfigure}
%\begin{subfigure}{0.33\textwidth}
%includegraphics[width=0.9\linewidth]{omega_0.35.pdf}
%\end{subfigure}
\caption{{\color{black}Transfer probability, $|\alpha(t\to\infty)|^2$, as a function of the electric field, $E$ (in the units of $V/a$). The markers show the numerically exact solution of Eq.~(\ref{eq:system_of_eqs}) with  $\mu=0$. The solid curves are the corresponding Landau-Dykhne results. The left panel is for $\hbar\omega=0.21$eV, the right panel is for $\hbar\omega=0.31$eV. }}
\label{fig:comparison_numerics1}
\end{figure}

\begin{figure}
\includegraphics[scale=0.7]{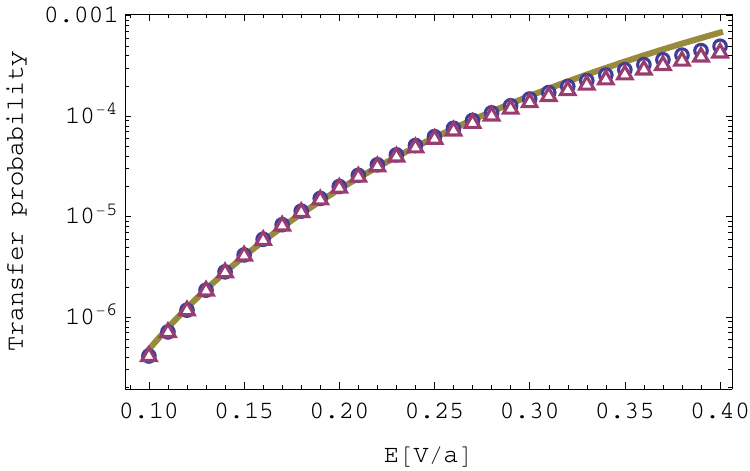}
\caption{{\color{black}Transfer probability, $|\alpha(t\to\infty)|^2$, as a function of the electric field, $E$ (in the units of $V/a$) in the presence of a spin-electric term. The markers show the numerically exact solution of Eq.~(\ref{eq:system_of_eqs}) with (triangles) and without (circles) the spin-electric term. For completeness, we also show the corresponding Landau-Dykhne results with $\mu=0$ (see the right panel of Fig.~\ref{fig:comparison_numerics2}).}}
\label{fig:comparison_numerics2}
\end{figure}

Having confirmed the validity of the Landau-Dykhne approximation for our parameters, we now investigate the effect of the spin-electric term. To this end, we solve Eq.~(\ref{eq:system_of_eqs}) numerically with $\mu=0.3 ea$~\cite{Volosniev2023a}. We present our results in Fig.~\ref{fig:comparison_numerics2}. We see only a minor effect of the spin-electric term, which validates the expressions presented in the main text.

} 
\vspace{1em}

\noindent {\bf Ionization in the presence of a frozen-in electric field.}

\vspace{1em}

\noindent Let us assume that the vector potential has the form
\begin{equation}
\mathbf{A}=-\mathbf{E}_{DC}(t-t_i)-\mathbf{E}_{AC}\frac{\sin(\omega t)}{\omega},
\end{equation}
where $\mathbf{E}_{DC}$ and $\mathbf{E}_{AC}$ are constant vectors that define the strength of the DC and AC electric fields, respectively; $t_i$ is some constant that is determined by initial conditions. We assume that $\mathbf{A}(t=0)=0$, which means that the difference between energy levels at $t=0$ is $\Delta$. This implies that $t_i=0$. 

It is clear that to satisfy $\mathcal{E}_{f}(t)=\mathcal{E}_{i}(t)$, $t$ should be imaginary, i.e., $t=i\tau$, where $\tau$ satisfies the equation
\begin{equation}
 E_{DC}^2\tau^2+2(\mathbf{E}_{DC}\mathbf{E}_{AC})\frac{\tau\sinh(\omega \tau)}{\omega}+
E_{AC}^2\frac{\sinh^2(\omega \tau)}{\omega^2} = \frac{\Delta^2}{(e a)^2(\Delta t_3+16 t^2)}.
\end{equation}
We assume that 
$E_{DC}\to 0$ so that 
\begin{equation}
 e a E_{DC} \tau\cos(\theta)+
e a E_{AC}\frac{\sinh(\omega \tau)}{\omega} \simeq  \sqrt{\frac{\Delta^2}{\Delta t_3+16 t^2}},
\end{equation}
where $\theta$ is the angle between $\mathbf{E}_{AC}$ and $\mathbf{E}_{DC}$.
We can solve this equation iteratively. At the first iteration, we neglect the static field completely: 
\begin{align}
e a E_{AC}\frac{\sinh(\omega \tau^0)}{\omega} \simeq  \sqrt{\frac{\Delta^2}{\Delta t_3+16 t^2}} \to \tau^0 = \frac{1}{\omega}\mathrm{arsinh}\left(\frac{\omega}{eaE_{AC}} \sqrt{\frac{\Delta^2}{\Delta t_3+16 t^2}}\right). 
\end{align}
This expression reproduces the result for one-color tunnel ionization. At the second iteration, we derive:
\begin{align}
 e a E_{DC} \tau^0\cos(\theta)+
e a E_{AC}\frac{\sinh(\omega \tau^1)}{\omega} \simeq  \sqrt{\frac{\Delta^2}{\Delta t_3+16 t^2}} \to \\
\tau^1=\frac{1}{\omega}\mathrm{arsinh}\left(\frac{\omega}{eaE_{AC}} \sqrt{\frac{\Delta^2}{\Delta t_3+16 t^2}}-\frac{ E_{DC}}{E_{AC}} \omega \tau^0\cos(\theta)\right)\nonumber.
\end{align}
We have checked numerically that for $E_{DC}\ll E_{AC}$, $\tau^1$ approximates $\tau$ well. 

The expression for $\tau_1$ allows us to compute $W_{fi}$:
\begin{equation}
W_{fi}\sim \mathrm{exp}\left(-2\mathrm{Im}\int_0^{i\tau^1} \left[\Delta+\frac{\Delta t_3 + 16 t^2}{2\Delta}\left(ea \mathbf{A}\right)^2\right]\mathrm{dt}\right),
\end{equation}
so that 
\begin{equation}
W_{fi}\sim e^{-2\tau^1 \Delta 
+(ea E_{AC})^2\frac{\Delta t_3 + 16 t^2}{4\Delta} \left[\frac{\sinh(2\omega \tau_1)-2\omega\tau_1}{\omega^3} + \frac{8E_{DC}}{E_{AC}}\cos(\theta)\frac{\omega \tau_0\cosh(\omega\tau_0)-\sinh(\omega \tau_0)}{\omega^3}\right]},
\end{equation}
where, for consistency of derivations, we use $\tau_0$ when we multiply by $E_{DC}$. Note that because the direction of $\mathbf{E}_{AC}$ is defined only up to a phase $\pi$, we should use absolute value of $\cos(\theta)$ in the expression.
 
 \vspace{1em}

\noindent {\bf Additional information in support of the bi-molecular-recombination origin of PL. }

\vspace{1em}
 
\noindent {\color{black} Assuming that the origin of PL is bi-molecular recombination, we conclude that $I_{PL}\sim W^2 $ (we shall omit the subscript $``fi"$ for simplicity) for a dilute population of charge carriers. This theoretical prediction describes our experimental data quantitatively for `weak' values of the electric field across two orders of the magnitude of the PL signal and for different frequencies, see Figs.~\ref{fig:fig_app_4um} and~\ref{fig:fig_app_2um}. Figure~\ref{fig:fig_app_4um} shows the PL signal from Fig.~1D of the main text together with the prediction based upon the Landau-Dykhne adiabatic approximation, $W^2$. The theoretical result does not have any fit parameters except the overall prefactor, which is beyond this work. [Indeed, it cannot be calculated using the Landau-Dykhne approach and also not measurable experimentally.] Given the fact that ionization rate is exponentially sensitive to the material properties, the quantitative agreement in Fig.~\ref{fig:fig_app_4um} is highly non-trivial. Figure~\ref{fig:fig_app_4um} also shows possible theoretical curves for multiphoton ionization~\cite{Nathan1985}, $E_{AC}^{32}$ (assuming bi-molecular recombination) and $E_{AC}^{16}$ (assuming mono-molecular recombination).

\begin{figure}
\includegraphics[scale=0.25]{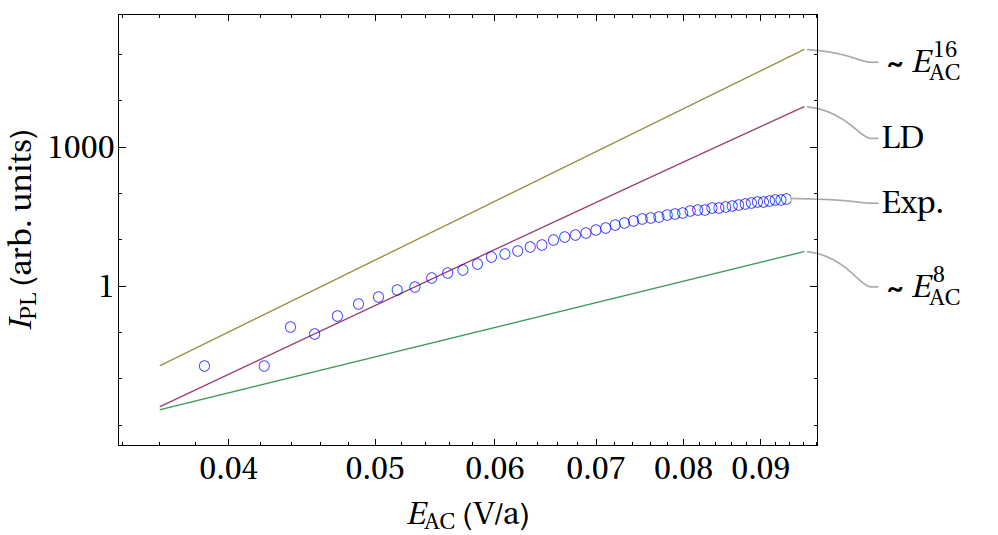}
\caption{{\color{black}Experimental PL spectra under $\lambda=2\mu$m pumping of a single-crystal MAPbBr$_3$, together with our theoretical prediction based upon the Landau-Dykhne (LD) adiabatic approximation (see the text for details). We also present power-law dependencies $E_{AC}^{16}$ and $E_{AC}^{8}$.}}
\label{fig:fig_app_2um}
\end{figure}

\begin{figure}
\includegraphics[scale=0.23]{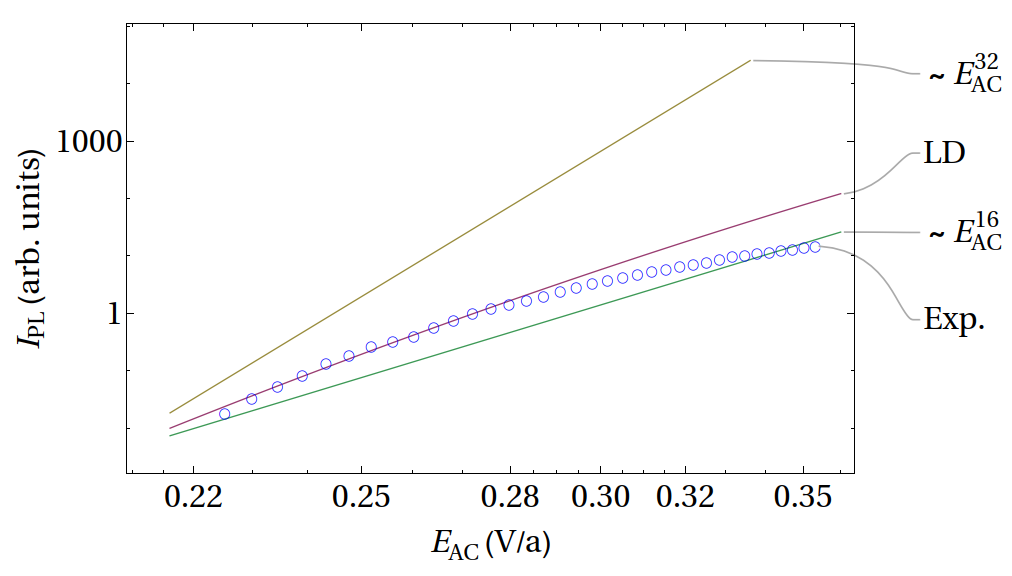}
\caption{{\color{black}Experimental PL spectra under $\lambda=4\mu$m pumping of a single-crystal MAPbBr$_3$, together with our theoretical prediction based upon the Landau-Dykhne (LD) adiabatic approximation (see the text for details). We also present power-law dependencies $E_{AC}^{32}$ and $E_{AC}^{16}$.}}
\label{fig:fig_app_4um}
\end{figure}

Figure~\ref{fig:fig_app_2um} demonstrates the PL signal for a $2\mu$m driving field. Here, the PL signal appears at smaller values of the electric field, which together with larger photon energy ($\simeq 0.6 $eV) translates into a large value of the Keldysh parameter $\gamma_K\simeq 5$. This implies a multi-photon nature of ionization in this case, which we confirm by comparing to power-law dependencies $E_{AC}^{16}$ (assuming bi-molecular recombination) and $E_{AC}^{8}$ (assuming mono-molecular recombination). 
Two remarks are in order here. First, we see that the power law $E_{AC}^{16}$ describes the data for a $2\mu$m driving field well. This reinforces the hypothesis of the bi-molecular origin of PL in our experiment. Second, the fact that the power law $E_{AC}^{32}$ describes the data for a $4\mu$m driving field poorly confirms our interpretation of the PL signal in terms of the diabatic tunnelling ionization.  
    }

\vspace{3em}
 
\noindent {\bf Additional information for the discussion on frozen-in electric fields}

\vspace{1em}

\noindent As is mentioned in the main text, the two-fold symmetric polarization dependence in Fig.~2B of the main text appears surprising since the measurement is performed at room temperature where MAPbBr$_3$ is expected to be in the cubic phase. To further understand this, we approximate the rate of tunneling ionization with the following expression
\begin{equation}
W_{fi}\sim e^{-\frac{5}{3}\frac{\Delta}{eaE}\sqrt{\frac{\Delta^2}{\Delta t_3 + 16 t^2}}},
\end{equation}
which is valid in an adiabatic regime. We see that there are a few possibilities to change the rate as a function of the angle. 
For example, either $E$ or $t_3$ and $t$ can depend on the angle. Let us first assume that $E$ is independent of the angle and that $t\to t+\delta t$ and $t_3\to t_3 + \delta t_3$, where 
$\delta t$ and $\delta t_3$ determine the change of the hopping integrals.
 In this case, for the parameters of our system, we derive
\begin{equation}
W_{fi}\sim e^{-1.4\frac{\Delta}{eaE}}e^{0.5  \frac{\delta}{eaE}},
\end{equation}
where $\delta=\delta t_3+32 t\frac{\delta t}{\Delta}\simeq \delta t_3 + 8 \delta t$. Assuming that $\delta t_3\sim \delta t $, we have $\delta \simeq 10 \delta t$. Our data show that for $eaE=0.140$eV the change in PL intensity can be as large as 40, see Fig.~2B of the main text. 
This means that the change of the hopping coefficient can be $\delta t\simeq 0.05$eV. Such a change implies that the effective mass is modified by about $10-20$\%, which can be achieved if the intrinsic strains in a single-crystal LHPs are of the order of 0.5\%~\cite{Chen2020}. Depending on the preparation of the LHP single crystal, it can either be strain free or have strains $\simeq 0.1$\%~\cite{Li2024}. We are not aware of a mechanism that can lead to a large renormalization of the effective mass under such weak strains. Therefore, we conclude that the electric field $E$ depends on angle, which 
leads to the hypothesis of ferroelectric domains presented in the main text.

\begin{figure*}[t]
\includegraphics[scale=0.4]{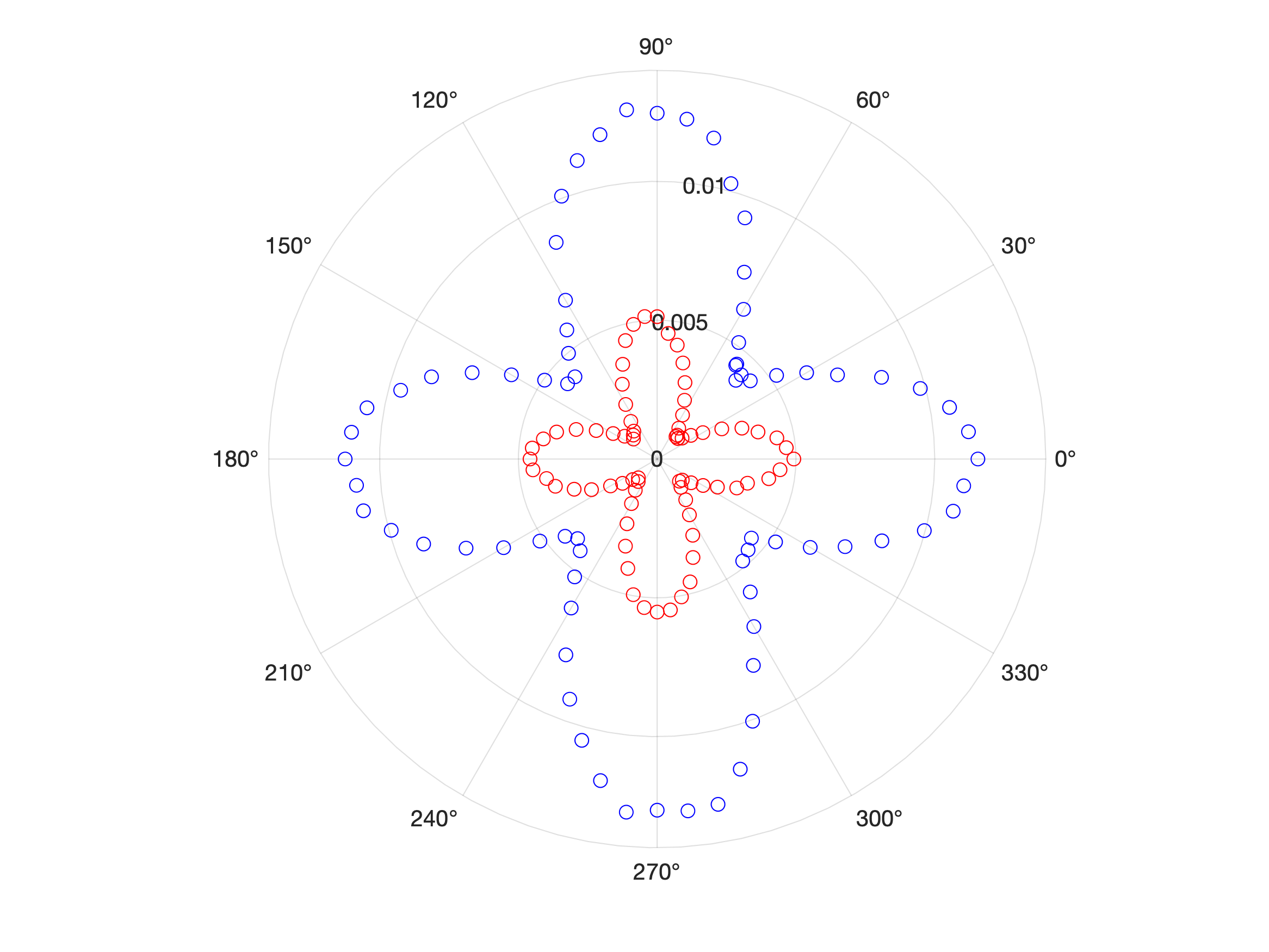}
\caption{Polarization scan taken at $E_{AC}\simeq 0.284$V/a (red markers) and $E_{AC}\simeq 0.312$V/a (blue markers). The PL intensity is shown in arbitrary units. }
\label{fig:fig_SM_four_leaf}
\end{figure*}

In Fig.~\ref{fig:fig_SM_four_leaf} we present a PL scan with a focal spot diameter of 126$\mu$m (1/$e^2$).

%%%%%%%%%%%%%%%%%%%%%%%%%%%%%%%%%%%
%%%%%%%%%%%%%%%%%%%%%%%%%%%%%%%%%%%

\vspace{3em}
\newpage

\noindent \textbf{Hysteresis of PL intensity }

\vspace{1em}
\noindent The samples show limited hysteresis with respect to the intensity scaling. As an example we show the data for MAPbBr$_3$ obtained at 300K for a pump wavelength of 4$\mu$m.
Note that in all our intensity dependent studies care was taken not to exceed values that would lead to substantial hysteresis. For the given example, this would mean an upper limit of 8$\times10^{14}$W/m$^2$ on the applied intensities.

\begin{figure}
\includegraphics[scale=0.5]{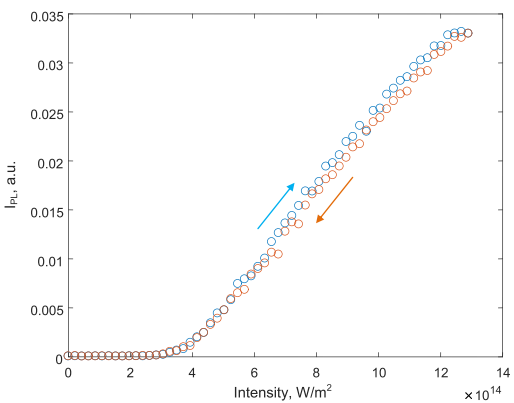}
\caption{{Hysteresis of the PL intensity for MAPbBr$_3$ at a pump wavelength of 4$\mu$m.}}
\label{fig:figSupp3}
\end{figure}

\vspace{3em}

\noindent {\bf Abraded samples of MAPbBr$_3$}

\vspace{1em}

\begin{figure}
\includegraphics[scale=0.1]{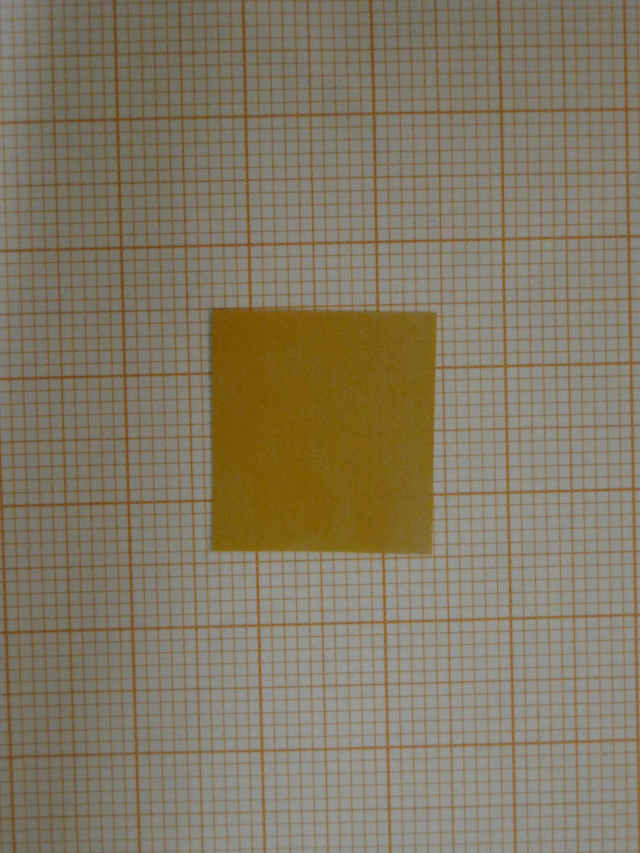}
\caption{Abraded sample}
\label{fig:figSupp1}
\end{figure}

\noindent For a two-color experiment, we mechanically grind the sample by rubbing an intact single crystal of MAPbBr$_3$ against a fine 1 {\textmu}m-grit Al$_2$O$_3$. As a result, we obtain a sample consisting of a substrate (sandpaper) covered by a powder of perovskite microcrystals as shown in Fig.~\ref{fig:figSupp1}. Our experience shows that microcrystal samples obtained this way are very robust, retaining the ability to PL under infrared-irradiation for years after being manufactured even if stored in atmosphere. The additional advantage of such samples as compared to single-crystal ones is the former being more sensitive than the latter since local stresses as a rule tend to enhance PL efficiency.

\begin{figure}
\includegraphics[width=0.85\textwidth]{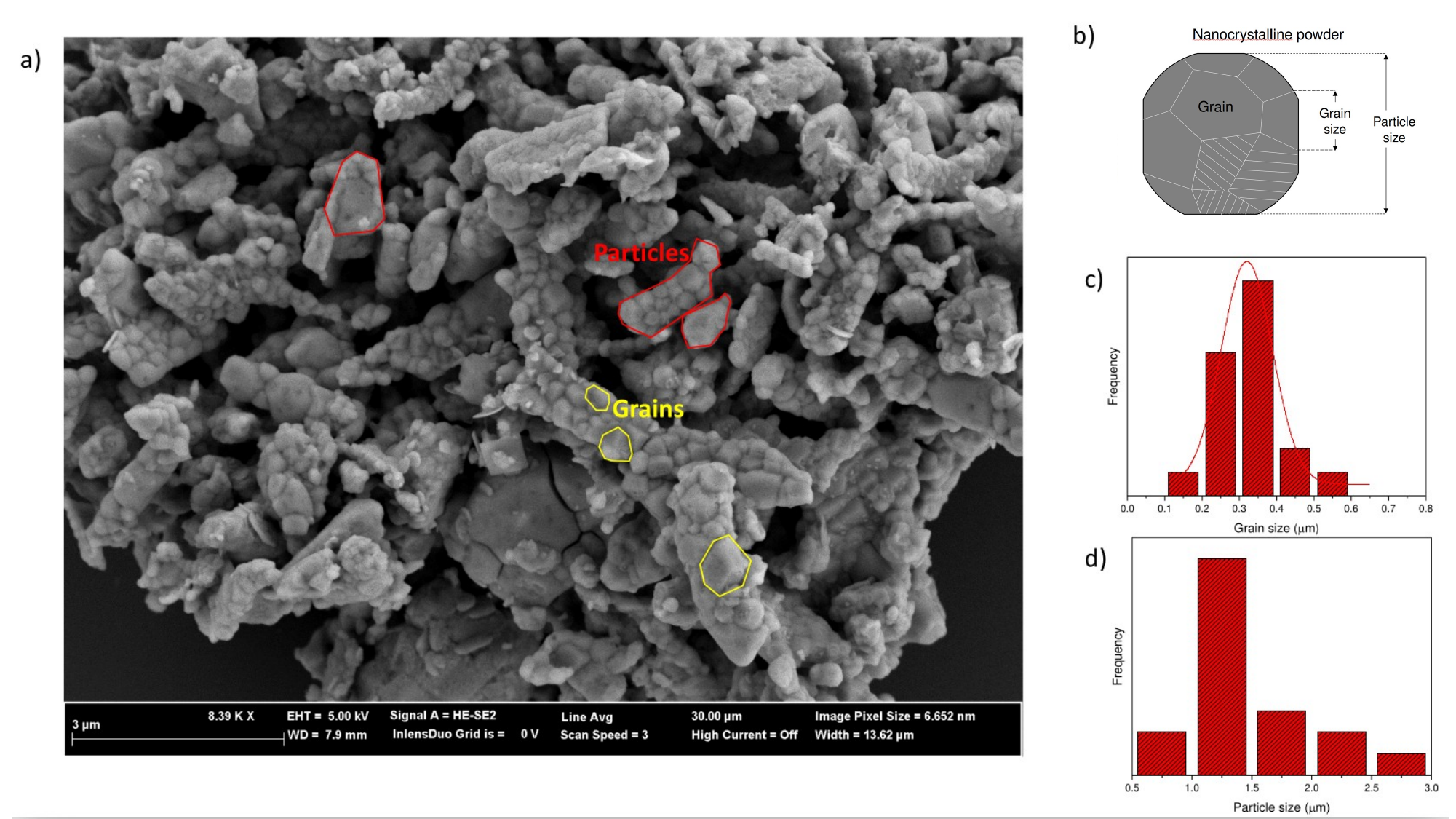}
\caption{a) Abraded sample as imaged by a SEM microscope with `particles' and `grains' highlighted in red and yellow, respectively; b) a schematic representation of a grain;  c) grain size; d) particle size distributions.}
\label{fig:figSupp2}
\end{figure}

To characterize the morphology of the microcrystal comprising the abraded samples, we first cut the sandpaper with microcrystals on top of them into small mm-sized pieces, which were then dropped into hexane and sonicated for 5 minutes. The resulting solution was drop casted on silicon substrates and allowed to dry. In this way prepared microcrystal samples were subsequently imaged by a SEM microscope. The observed grain sizes are shown in Fig.~\ref{fig:figSupp2}. From the evaluation of the grain/particle statistics, we obtain average grain size of about 320 nm $\pm$ 60 nm and a particle size of 1.3 $\mu$m $\pm$ 0.23 $\mu$m.

To investigate cooperation of dynamic fields and AC-biasing, the sample is placed in a defocused 1030 nm beam while being kept in the focus of the 4500 nm pump beam. Images of the resulting PL were taken using a CMOS camera (Point Grey Research PGR-CM3-U3-50S5M-CS).

\vspace{3em}

\noindent {\bf Additional data for a two-color experiment}

\vspace{1em}

\begin{figure}
\includegraphics[width=0.7\textwidth]{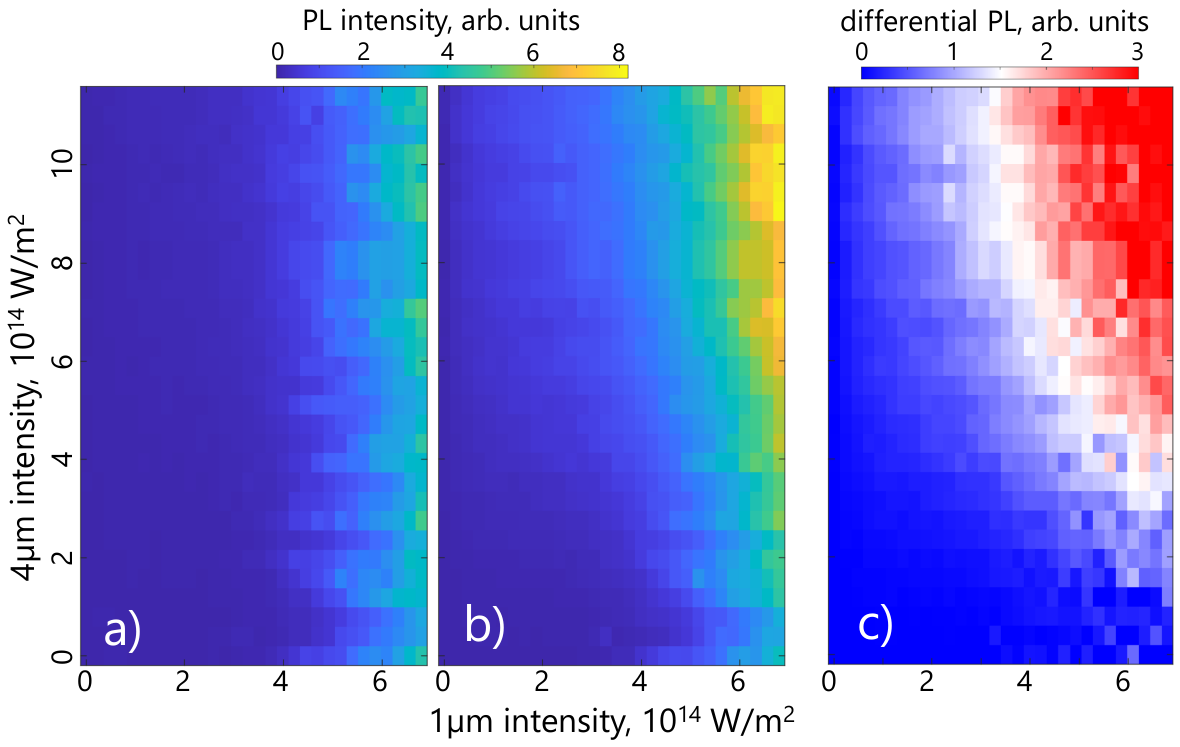}
\caption{{\color{black}Two-color experiment on a single-crystal MAPbBr$_3$: cumulative PL as a function of the intensities of 1$\mu$m and 4$\mu$m with $\delta t= 7$ps  (``off'' position, panel a) and overlapped in time (``on'' position, panel b) and the differential PL defined as the difference between ``on''- and ``off''-positions (panel c);}}
\label{fig:figSupp_resub}
\end{figure}

\noindent {\color{black} As a further illustration of the two-color experiment on a single-crystal MAPbBr$_3$, we present in  Fig.~\ref{fig:figSupp_resub} the 2D dependence of PL on the intensity of each infrared beam.}

 %%%%%%%%%%%%%%%%%%%%%%%%%%%%%%%
\begin{figure}[h]
\includegraphics[width=0.5\textwidth]{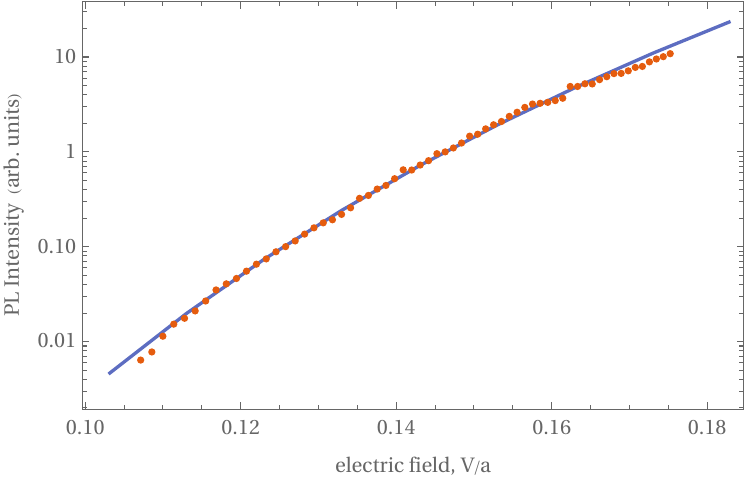}
\caption{ PL intensity for CsPbBr$_3$ under 4$\mu$m pumping (markers).  The solid curve demonstrates the functional dependence of $(W_{fi})^{1.2}$.}
\label{fig:fig2_SM}
\end{figure}
%%%%%%%%%%%%%%%%%%%%%%%%%%%%%%%%%

\vspace{3em}
 
\noindent {\bf Photoluminescence from CsPbBr$_3$}

\vspace{1em}

\noindent To demonstrate tunnel ionization in another perovskite, we study the PL spectra of CsPbBr$_3$ irradiated by a 4 {\textmu}m laser, see Fig.~\ref{fig:fig2_SM}. As for MAPbBr$_3$, we observe that the PL intensity is exponentially sensitive to the strength of the external electric field. The crucial difference however is that our analytical predictions  for MAPbBr$_3$ do not describe the observed data in CsPbBr$_3$. This could have been anticipated as in our derivations we relied on the cubic structure of the material. However, CsPbBr$_3$ (unlike MAPbBr$_3$) has an orthorhombic structure at room temperature so a more careful analysis must be in order. Further, we note that if the density of charge carriers is small (note that the intensity of the external electric field is relatively small in our experiment with CsPbBr$_3$), then an electron can recombine only with its partner hole created in the tunnel ionization process or with localized trap states. These processes are not included in bimolecular radiative recombination used in our analysis of MAPbBr$_3$.

 To quantify the deviation from the $W_{fi}^2$-behavior observed in MAPbBr$_3$ at small fields, we fit our CsPbBr$_3$ data  with $W_{fi}^\delta$, where $\delta$ is a fitting parameter. We find that $\delta\simeq 1.2$ can be used to describe the data. As this value is close to unity, we hypothesize that the radiative processes in CsPbBr$_3$ for the considered parameters are mainly of monomolecular origin, see also Refs.~\cite{Abiedh2021,Zhang2024}.

\bibliography{ref}

\end{document}